# Physical modeling of the soil swelling curve vs. the shrinkage curve

V.Y. Chertkov*

Division of Environmental, Water, and Agricultural Engineering, Faculty of Civil and Environmental Engineering, Technion, Haifa 32000, Israel

**Abstract.** Physical understanding of the links between soil swelling, texture, structure, cracking, and sample size is of great interest for the physical understanding of many processes in the soil-air-water system and for applications in civil, agricultural, and environmental engineering. The background of this work is an available chain of interconnected physical shrinkage curve models for clay, intra-aggregate matrix, aggregated soil without cracks, and soil with cracks. The objective of the work is to generalize these models to the case of swelling, and to construct the physical-swelling-model chain with a step-by-step transition from clay to aggregated soil with cracks. The generalization is based on thorough accounting for the analogies and differences between shrinkage and swelling and the corresponding use, modification, or replacement of the soil shrinkage features. Two specific soil swelling features to be used are: (i) air entrapping in pores of the contributing clay; and (ii) aggregate destruction with the formation of new aggregate surfaces. The input for the prediction of the swelling curve of an aggregated soil coincides with that of the available model of the shrinkage curve. The analysis of available data on the maximum shrink-swell cycle of two soils with different texture and structure, accounting for sample size is conducted as applied to swelling curves and to the residual crack volume and maximum-swelling-volume decrease after the shrink-swell cycle. Results of the analysis show evidence in favor of the swelling model chain.

*Corresponding author. Tel.: 972-4829-2601.
*E-mail address:* agvictor@tx.technion.ac.il; vychert@ymail.com (V.Y. Chertkov).

**1. Introduction**

The number of published works devoted to soil swelling is small compared to that of works devoted to soil shrinkage, although in natural conditions the stages of soil shrinkage and swelling alternate. Physical understanding of the links between soil swelling, texture, structure, cracking, and sample size is of great interest for both the physical understanding of many processes in the soil-air-water system and for applications in civil, agricultural, and environmental engineering. Indeed, soil shrinkage is accompanied by cracking that strongly influences the hydraulic and other properties of the soil. The swelling that follows does not usually lead to the total closing of the cracks, and the actual observed soil cracking (residual cracking) depends on both shrinkage and swelling. In addition, a key phenomenon such as the experimentally observed decrease of the maximum swelling volume of an aggregated soil after a shrink-swell cycle in the maximum possible water content range (see e.g., [1]), requires a physical quantitative explanation. It is noteworthy that works devoted to soil swelling usually give the data on swelling volume as a function of wetting time (e.g., [2,3]). We could find only one work [1] with experimental data on the swelling curve as the dependence of the void ratio on the moisture ratio like the usual presentation of a shrinkage curve. Such a presentation of the swelling curve (or that of the specific soil volume vs. gravimetric water content) is important for the following comparison between the shrinkage and swelling curves and considering the results of the joint action of shrink-swell stages. The physical models of the swelling curve are currently lacking. By this, we mean the models deriving the swelling curve from the soil texture and structure, rather than just fitting some mathematical approximations to the experimental data as, e.g., [1].

Our *general objective* in this work is to suggest some physical approach to constructing the swelling curve of aggregated soils in the maximum possible water content range between the zero water content and maximum swelling point when swelling follows previous shrinkage (without loading). The general methodology of the approach is reduced to constructing the swelling curve as the generalization of earlier proposed and validated shrinkage models, accounting for both similarities and differences of shrinkage and swelling. At present a number of soil shrinkage curve models are available that are based on fitting of several parameters, different for different models [4-12], as well as the physical models [13-17]. The latter models are subject to the generalization (see below). The chain of the interconnected physical shrinkage models (in the maximum possible water content range) that are basic in the generalization relies on the concepts of the inter- and intra-aggregate structure of inorganic soils presented in Fig.1. The intra-aggregate structure includes two basic peculiarities: the deformable, but non-shrinking and non-swelling aggregate surface layer (or interface layer) [13-15] and relatively large lacunar pores in the clay matrix that contributes to aggregates [18, 14]. The first in the above chain, the model of the clay shrinkage curve [19,20], is based on a simple physical approximation (see Section 2.1). The second model of the shrinkage curve of an intra-aggregate matrix without [13,16] and with [14] lacunar pores (Fig.1) is obtained as a result of transition from the above pure clay model. The third model of the reference shrinkage curve [13-15] (i.e., the case of sufficiently small samples without cracks) is obtained as a result of transition from the second model. Finally, the fourth model of the shrinkage curve of a sample or layer of any size and with cracks [17] is obtained as a result of transition from the reference shrinkage curve model. Relying on the above shrinkage model chain we should consider the corresponding generalization to the swelling case (for the specification of the swelling conditions see Section 2.2): (i) for each link of the chain; and (ii) for each transition from link to link. That is, for each link we should consider the corresponding swelling "twin" and formulate the corresponding swelling model that combines the analogies based on the shrinkage case, where it is possible, and the new features relevant to the swelling case starting from the zero water content. Two specific soil swelling features to be used are: (i) air entrapping in pores of the dry contributing clay; and (ii) aggregate destruction with the formation of new aggregate surfaces (the air entrapment with aggregate destruction have been noted in many works, e.g., among others [21,22,1]). In other words, relying on the above shrinkage model chain, we intend to construct the corresponding swelling model chain. Thereby, we intend to consider the maximum shrink-swell cycle of aggregated soil, starting from the maximum swelling state. Before construction of each swelling link, for the reader convenience and shortening the exposition we briefly give the major points of the corresponding shrinkage link we only rely on in the generalization to the swelling case. Eventually, we model the swelling curve of aggregated inorganic soils with cracks (in the maximum possible water content range), starting from the construction of the clay swelling curve, to the swelling curve of the intra-aggregate matrix, and reference swelling curve (i.e., without cracks). The primary checking and validating of the different aspects of the swelling model chain is conducted using the relevant available data [1] on the shrink-swell cycle of two inorganic soils with different texture and structure, accounting for sample size. Notation of the values that repeat in the text is summarized at the end of the paper.

## 2. The swelling curve of a pure disaggregated clay
### 2.1. The major points of an available model in the shrinkage case

As previously noted, heuristic considerations that enable the construction of the clay swelling curve are inspired by the available model of the shrinkage curve of a clay matrix (or clay paste) [19,20]. In the model the shrinkage curve is first found in relative coordinates, $v(\zeta)$ where $v$ is the ratio of clay volume to its maximum in the solid state (the liquid limit), and $\zeta$ is the ratio of water content to its maximum in the solid state. The single-valued expression of the clay shrinkage curve, $v(\zeta)$ flows out of the simple physical considerations that were experimentally based. (1) At drying in the water content range of the basic shrinkage, the increments of $v$ and $\zeta$ are proportional to each other (Fig.2, curve $v(\zeta)$)

$$v(\zeta)=v_s+(1-v_s)\zeta, \qquad \zeta_n \leq \zeta \leq \zeta_h \tag{1}$$

where $v_s$ is the relative volume of clay solids; $\zeta_h \cong 0.5$ is the maximum swelling point of the clay [13,16]; $\zeta_n$ is the end point of basic shrinkage (the air-entry point). (2) At drying in the water content range of the zero shrinkage (Fig.2, curve $v(\zeta)$)

$$v(\zeta)=v_z, \qquad 0 \leq \zeta \leq \zeta_z \tag{2}$$

where $\zeta=\zeta_z$ is the shrinkage limit; $v_z \equiv v(\zeta_z)$ is the relative volume of the oven-dried clay matrix. (3) The boundary values of the relative coordinates, $\zeta_z$ and $\zeta_n$ as well as $v_z$ and $v_n$ (Fig.2, curve $v(\zeta)$) for any clay satisfy the physical conditions

$$\zeta_n - \zeta_z \ll 1, \qquad v_n - v_z \ll 1. \tag{3}$$

For this reason, at drying in the small intermediate range, $\zeta_z \leq \zeta \leq \zeta_n$ one can present $v(\zeta)$ as an expansion in powers of $(\zeta - \zeta_z)$ and then be limited by the second power as

$$v(\zeta)=v_z+a(\zeta-\zeta_z)^2, \qquad \zeta_z \leq \zeta \leq \zeta_n. \tag{4}$$

(4) The saturation degree of the clay matrix at shrinkage, $F(\zeta)$ is linked with $\zeta$ and $v(\zeta)$ and, in particular, $F_z \equiv F(\zeta_z)$ determines $\zeta_z$ as [19]

$$F=(1-v_s)\zeta/(v-v_s), \qquad 0 \leq \zeta \leq \zeta_h \tag{5}$$

$$\zeta_z=(v_z-v_s)F_z/(1-v_s). \tag{6}$$

(5) Accounting for Eq.(6), the unknown values $\zeta_n$ and $a$ in Eq.(4) can be expressed through $v_s$, $v_z$, and $F_z$ from the two smoothness conditions of the clay shrinkage curve, $v(\zeta)$ at $\zeta=\zeta_n$ (Fig.2) to be

$$\zeta_n=(v_z-v_s)(2-F_z)/(1-v_s), \qquad a=(1-v_s)^2/[4(v_z-v_s)(1-F_z)]. \tag{7}$$

The two similar smoothness conditions of $v(\zeta)$ at $\zeta=\zeta_z$ (Fig.2) are automatically satisfied by the presentation of $v(\zeta)$ from Eqs.(2) and (4).
(6) The physical link between $F(\zeta)$ and the pore-size distribution of the clay matrix as well as the links between macro-parameters of the clay matrix, $v_s$, $v_z$ and characteristic sizes of clay particles (micro-parameters), enable the finding of $F_z$ as a function of $v_s$ and $v_z$, and, thereby, excluding the $F_z$ parameter from the shrinkage curve as an independent one [20]. As a result, the clay shrinkage curve is only expressed through two physical clay matrix parameters, $v_s$ and $v_z$. They are specific for a particular clay and reflect its mineralogy as well as the concentration and type of cations in the water. On the macro-level $v_s$ and $v_z$ have clear physical meaning (see above) and can be measured independently of the clay shrinkage curve [20,23]. (7) Finally, the transition from the relative coordinates ($\zeta,v$) to customary coordinates ($\overline{w},V$) (the specific volume, $V$ vs. gravimetric water content, $\overline{w}$ of the clay matrix) or ($\theta, e$) (the void ratio, $e$ vs. moisture ratio, $\theta$ of the clay matrix) is realized by [19,20]

$$V=v/(v_s\rho_s) \ , \qquad \overline{w}=((1-v_s)/v_s)(\rho_w/\rho_s)\zeta \tag{8}$$

$$e=v/v_s-1 \ , \qquad \theta=((1-v_s)/v_s)\zeta \tag{9}$$

where $\rho_w$ is the water density and $\rho_s$ is the density of the clay solids. Qualitatively, the $V(\overline{w})$ and $e(\theta)$ dependences are similar to $v(\zeta)$ in Fig.2. According to Eqs.(8) and (1) the characteristic slope, $dV/d\overline{w}$ of the clay shrinkage curve in coordinates ($\overline{w},V$) in the basic shrinkage range, $\overline{w}_n \leq \overline{w} \leq \overline{w}_h$ (corresponding to $\zeta_n \leq \zeta \leq \zeta_h$) does not depend on clay type. Indeed, in this range $dV/d\overline{w}=(dv/d\zeta)/[(1-v_s)\rho_w]=1/\rho_w$.

Besides the clay shrinkage curve, for the following transition from clay to soil shrinkage [13,14] and swelling (see section 4), it is also important to note the presentation of clay matrix porosity, $P(v(\zeta))$ at shrinkage [19]

$$P(v(\zeta))=1-v_s/v(\zeta) \ , \quad v_z\leq v\leq v_h \ , \quad 0\leq\zeta\leq\zeta_h \tag{10}$$

as well as the maximum and minimum internal sizes, $r_m(v(\zeta))$ and $r_o(v(\zeta))$, respectively, of clay matrix pores (excluding pore wall thickness) [19]

$$r_m(v(\zeta))=r_{mM}\,v(\zeta)^{1/3}(1-v_s/[Av(\zeta)]) \ , v_z\leq v\leq v_h \ , \quad 0\leq\zeta\leq\zeta_h \tag{11a}$$

$$r_o(v(\zeta))=r_{mM}v(\zeta)^{1/3}(\gamma v_s/A)(1-1/[\gamma v(\zeta)]), \ (r_o\to 0 \text{ at } v_z\to 0.11) \ v_z\leq v\leq v_h, 0\leq\zeta\leq\zeta_h \tag{11b}$$

where $r_{mM}$ is the maximum external size of the clay pores (including pore wall thickness) at $\zeta=1$; $A\cong 13.57$ and $\gamma\cong 9$ are the characteristic constants of the clay matrix.

## 2.2. Generalization to the swelling case

The clay (paste) swelling curve in relative coordinates will be denoted as $\hat{v}(\zeta)$ (unlike the shrinkage curve, $v(\zeta)$). Similar to the shrinkage curve (Fig.2; curve $v(\zeta)$) the clay swelling curve, $\hat{v}(\zeta)$ can be characterized by some general qualitative view (Fig.2; curve $\hat{v}(\zeta)$). Unlike the clay shrinkage curve, the swelling curve is presented by one curved line without linear sections. Similar to the shrinkage case the specific physical features of the curved line (see below) allow one to present the swelling curve by some simple single-valued expression (without fitting parameters). It is important to note that in the course of the generalization we use some results from [19,20] which in fact relate to both the shrinkage and swelling case (see below the results for the saturation degree, clay porosity, maximum and minimum internal size of clay pores, and transition from relative to customary coordinates). One can note the following physical features of the swelling curve (Fig.2; curve $\hat{v}(\zeta)$): (i) at $\zeta=0$ and $\zeta=\zeta_h$ the $\hat{v}(\zeta)$ curve has values, respectively, as

$$\hat{v}(0)=v_z \qquad \text{and} \qquad \hat{v}(\zeta_h)=v_h \ , \tag{12}$$

(ii) the slope of the $\hat{v}(\zeta)$ curve monotonously decreases at wetting from an initial maximum value at $\zeta=0$ (see below) to zero at $\zeta=\zeta_h$ as

$$d\hat{v}/d\zeta|_{\zeta=\zeta_h}=0 \ , \tag{13}$$

and (iii) the coordinates, $\zeta_h$, $v_h$, and $v_z$ satisfy the physical conditions (Fig.2; cf. Eq.(3))

$$\zeta_h \cong 0.5 < 1 \ , \qquad v_h - v_z << 1 \ . \tag{14}$$

One can present $\hat{v}(\zeta)$ in the range $0 \leq \zeta \leq \zeta_h$ (Fig.2) as an expansion in powers of ($\zeta-\zeta_h$). Then, in force of Eq.(14), it can be limited by some power, n of ($\zeta-\zeta_h$). After that one can take into account the conditions at $\zeta=\zeta_h$ from Eqs.(12) and (13). As a result $\hat{v}(\zeta)$ is presented to be

$$\hat{v}(\zeta)=v_h - b_2(\zeta-\zeta_h)^2 - \ldots - b_n(\zeta-\zeta_h)^n \ , \qquad 0 \leq \zeta \leq \zeta_h \ . \tag{15}$$

It is known that swelling conditions essentially influence the air entrapping and steepness of the swelling curve at small water contents (e.g., [21]). For this reason, in general, unlike the case of shrinkage (cf. Eq.(4)), at swelling n in Eq.(15) can be both equal and more than 2 reflecting the different swelling behavior of clay. The simplest case, n=2 leads to the slowest swelling. Below we consider namely this case assuming that it corresponds to the swelling conditions from [1] (core wetting by water vapor step by step, then saturation by capillary rise of water from the bottom). This assumption will be justified in Section 7.3. Then, using the condition at $\zeta=0$ from Eq.(12) and accounting for $v_h=0.5(1+v_s)$ (according to Eq.(1) and $\zeta_h \cong 0.5$), the unknown $b \equiv b_2$ value can be expressed through $v_s$ and $v_z$ as

$$b=(v_h-v_z)/\zeta_h^2=2(1+v_s-2v_z) \ . \tag{16}$$

It is worth noting that the initial (maximum) slope of $\hat{v}(\zeta)$ at $\zeta=0$ (Fig.2) being

$$d\hat{v}/d\zeta|_{\zeta=0}=2b\zeta_h=b=2(1+v_s-2v_z) \ , \tag{17}$$

can be more (as in Fig.2) and less than the maximum slope of the shrinkage curve, (1-$v_s$) (see Eq.(1)) in the basic shrinkage range (Fig.2) at the usual $v_s$ and $v_z$ values ($0.03 < v_s </\cong 0.2$ and $\max(v_s, 0.11) < v_z < v_h = 0.5(1+v_s) < 0.6$ [19,20]). In any case the sufficiently steep initial clay volume increase at wetting (Eq.(17)) is connected with air *entrapment* and compression in the part of the clay pore volume (e.g., [21,22,1]).

At the following wetting air is gradually displaced, and the swelling curve slope decreases to zero.

The derivation of the expression for the saturation degree at shrinkage, $F(\zeta)$ (Eq.(5)) [19] is not specific for the shrinkage case and also relates to the clay swelling curve, $\hat{v}(\zeta)$ if $v(0)=\hat{v}(0)$ and $v(\zeta_h)=\hat{v}(\zeta_h)$ (Fig.2). Therefore, denoting the saturation degree at swelling by $\hat{F}(\zeta)$ (unlike $F(\zeta)$ at shrinkage) we can write from Eq.(5)

$$\hat{F}(\zeta) = (1-v_s)\zeta/(\hat{v}(\zeta)-v_s) , \qquad 0\leq\zeta\leq\zeta_h . \qquad (18)$$

At maximum swelling (Fig.2; $\zeta=\zeta_h$ and $\hat{v}(\zeta_h)=v_h=v_s+(1-v_s)\zeta_h$) Eq.(18) gives $\hat{F}(\zeta_h)=1$ as it should be according to the physical meaning of $\hat{F}(\zeta_h)$. $F(\zeta_h)$ is also equal to unity. However, from Eq.(18) at $\zeta<\zeta_h$ $\hat{F}(\zeta)<1$. Unlike that $F(\zeta)=1$ at $\zeta_n\leq\zeta<\zeta_h$ (Fig.2).

Similar to the link between $F(\zeta)$ and the pore-size distribution of the clay matrix at shrinkage, there is the analogical link between $\hat{F}(\zeta)$ and the distribution at swelling. In case of shrinkage this link was used to find the $F_z(v_s, v_z)$ function and exclude the $F_z$ from the clay shrinkage curve, $v(\zeta)$ as an independent parameter [20]. In case of swelling we already have the expression for the swelling curve, $\hat{v}(\zeta)$ (Eqs.(15) and (16)) that is only determined by $v_s$ and $v_z$. Therefore, one can use the link between $\hat{F}(\zeta)$ and the pore-size distribution of the clay matrix for other purposes (see the use of Eq.(18) in Eq.(69)). Similar to the saturation degree case (Eqs.(5) and (18)), the expressions for the clay matrix porosity at swelling, $\hat{P}(\zeta)=P(\hat{v}(\zeta))$ and the maximum and minimum internal sizes of clay matrix pores at swelling, $\hat{r}_m(\zeta)=r_m(\hat{v}(\zeta))$ and $\hat{r}_o(\zeta)=r_o(\hat{v}(\zeta))$, that will be needed below, are obtained from Eqs.(10), (11a) and (11b), respectively, after replacement, $v(\zeta)\to\hat{v}(\zeta)$. Finally, note that transition from the swelling curve in the $(\zeta,\hat{v})$ coordinates to that in the $(\overline{w},\hat{V})$ or $(\theta,\hat{e})$ coordinates ($\hat{V}$ and $\hat{e}$ are the specific clay volume and void ratio, respectively, at wetting) is realized as usual by Eqs.(8) and (9) [19,20] after replacements: $v\to\hat{v}$, $V\to\hat{V}$, and $e\to\hat{e}$. The $\hat{V}(\overline{w})$ and $\hat{e}(\theta)$ dependences are qualitatively similar to $\hat{v}(\zeta)$ in Fig.2. According to Eqs.(8) (with $\hat{V}$ and $\hat{v}$) and (17) the characteristic *initial* slope, $d\hat{V}/d\overline{w}$ of the clay swelling curve depends on clay type (unlike the characteristic slope of the clay shrinkage curve in the basic shrinkage range; see point 7 of section 2.1). Indeed,

$$d\hat{V}/d\overline{w}|_{\overline{w}=0} = (d\hat{v}/d\zeta|_{\zeta=0})/[(1-v_s)\rho_w] = 2(1+v_s-2v_z)/[(1-v_s)\rho_w] . \qquad (19)$$

### 3. The swelling curve of the intra-aggregate matrix of a soil
#### *3.1. The major points of an available model in the shrinkage case*
(1) Besides clay, the intra-aggregate matrix (Fig.1) includes silt and sand grains as well as lacunar pores. Similar to the definition of the relative coordinates for clay, the relative volume, $u$ of the soil intra-aggregate matrix is the ratio of the matrix volume to its maximum in the solid state (at the liquid limit); the relative water content, $\zeta$ of the intra-aggregate matrix is the ratio of the matrix water content to its maximum in the solid state, and coincides with $\zeta$ for clay (see section 2.1) [13,14]. At any possible soil clay content, $0<c<1$, the relative volume of the intra-aggregate matrix, $u$ is linked with the relative volume of the contributing clay, $v$ as [13,14]

$$v(\zeta)=(u(\zeta)-u_{lp}(\zeta)-u_S)/(1-u_S), \tag{20a}$$

$$v_s=(u_s-u_S)/(1-u_S), \qquad v_z=(u_z-u_{lpz}-u_S)/(1-u_S), \qquad v_h=(u_h-u_{lph}-u_S)/(1-u_S), \tag{20b}$$

where $u_s$ and $u_S$ are the relative volume of the solid phase of the intra-aggregate matrix (silt and sand grains and clay particles) and the similar relative volume of the non-clay solids, respectively; $u_{lp}(\zeta)$ is the relative volume of the lacunar pores [14,15], and correspondingly, $u_z=u(\zeta_z)$, $u_h=u(\zeta_h)$, $u_{lpz}=u_{lp}(\zeta_z)$, $u_{lph}=u_{lp}(\zeta_h)$. The ways of estimating $u_s$, $u_S$, $u_z$, $u_h$, $u_{lpz}$, and $u_{lph}$ for a soil were considered in detail [13,14,17]. (2) The replacement of $v(\zeta)$, $v_z$, $v_h$, and $v_s$ in Eqs.(1), (2), (4)-(7) with $v(\zeta)$, $v_z$, $v_h$, and $v_s$ from Eqs.(20a)-(20b) leads to the shrinkage curve, $u(\zeta)$ of the intra-aggregate matrix (Fig.1) in relative coordinates as

$$u(\zeta) = \begin{cases} u_z, & 0 \leq \zeta \leq \zeta_z \\ u_z + u_{lp}(\zeta) - u_{lpz} + \dfrac{(1-u_s)^2}{4(u_z - u_s - u_{lpz})(1-F_z)}(\zeta-\zeta_z)^2, & \zeta_z < \zeta \leq \zeta_n \\ u_s + u_{lp}(\zeta) + (1-u_s)\zeta, & \zeta_n < \zeta \leq \zeta_h. \end{cases} \tag{21}$$

(3) Introduction of the lacunar factor, $k$ as the fraction of the clay pore volume decrease, $-du_{cp}$ that is equal to the lacunar pore volume increase, $du_{lp}$ [14], i.e.,

$$du_{lp}=-kdu_{cp}, \qquad 0\leq\zeta<\zeta_h \tag{22}$$

enabled one to find the $u_{lp}(\zeta)$ dependence entering Eq.(21) to be

$$u_{lp}(\zeta)=u_{lpz}-k(1-u_S)(v(\zeta)-v_z), \qquad 0<\zeta\leq\zeta_h. \tag{23}$$

(4) The lacunar factor, $k$ as a function of clay content, clay type, and soil texture was recently considered [24] to be

$$k(c/c_*)=[1-(c/c_*)^3]^{1/3}, \quad 0<c/c_*<1 \tag{24a}$$

$$k(c/c_*)=0, \quad 1<c/c_*<1/c_* \tag{24b}$$

where $c_*$ is the critical clay content [13] ($p$ is the porosity of the contributive silt and sand grains when they are in the state of *imagined* contact)

$$c_*=[1+(v_z/v_s)(1/p-1)]^{-1}. \tag{25}$$

(5) The qualitative view of $u(\zeta)$ from Eq.(21) is similar to $v(\zeta)$ (Fig.2). However, the slope, $du/d\zeta$ in the basic shrinkage range ($\zeta_n\leq\zeta\leq\zeta_h$) is [14]

$$du/d\zeta = (1-k(c/c_*))(1-u_s) , \qquad \zeta_n \leq \zeta \leq \zeta_h \qquad (26)$$

unlike $dv/d\zeta = (1-v_s)$. That is, the slope, $du/d\zeta$ depends (through $k(c/c_*)$) not only on the clay type ($v_s$, $v_z$), but also on clay content ($c$), and soil texture ($p$) if $c<c_*$ and $0<k<1$.
(6) The transition from the relative coordinates ($\zeta$, $u$) to customary ones ($w$, $U$) (specific volume vs. gravimetric water content of the soil intra-aggregate matrix) occurs in the usual way [13,14] as

$$w = ((1-u_s)/u_s)(\rho_w/\rho_s)\zeta , \qquad U = u/(u_s\rho_s) , \qquad 0 \leq w \leq w_h \qquad (w = c\overline{w}) \qquad (27)$$

where $w_h = w(\zeta_h = 0.5)$; $\rho_s$ being the (mean) density of solids (silt and sand grains and clay particles). The $U(w)$ specific volume is qualitatively similar to $V(\overline{w})$ and $v(\zeta)$ (Fig.2; see also dash curve 1 in Fig.3). However, the slope, $dU/dw$ of the shrinkage curve of the soil intra-aggregate matrix in the basic shrinkage range,

$$dU/dw = (1-k(c/c_*))/\rho_w , \qquad w_n \leq w \leq w_h \qquad (28)$$

is less than $dV/d\overline{w} = 1/\rho_w$ at $\overline{w}_n \leq \overline{w} \leq \overline{w}_h$ if $c<c_*$ and $0<k<1$, and depends (through $k(c/c_*)$; see Eqs.(24a) and (24b)) on clay type and content as well as soil texture.

### *3.2. Generalization to the swelling case*

Unlike the shrinkage curve, $u(\zeta)$, we denote the relative volume of the soil intra-aggregate matrix at swelling by $\hat{u}(\zeta)$ (similar to designations, $v$ and $\hat{v}$ for the relative volume of clay at shrinkage and swelling, respectively). Then, the transformations $v \to u$ (Eqs.(20a)-(20b)) at transitions from the contributing clay to soil intra-aggregate matrix at shrinkage, are totally retained in case of swelling and converted into transformations $\hat{v} \to \hat{u}$ after the symbol replacement, $v \to \hat{v}$ and $u \to \hat{u}$ to be

$$v(\zeta) \to \hat{v}(\zeta) , \qquad v_z \to \hat{v}_z = v_z , \qquad v_h \to \hat{v}_h = v_h , \qquad v_s \to \hat{v}_s = v_s , \qquad (29a)$$

$$u(\zeta) \to \hat{u}(\zeta), \qquad u_z \to \hat{u}_z = u_z , \qquad u_h \to \hat{u}_h = u_h , \qquad u_s \to \hat{u}_s = u_s , \qquad u_S \to \hat{u}_S = u_S , \qquad (29b)$$

$$u_{lp}(\zeta) \to \hat{u}_{lp}(\zeta) , \qquad u_{lpz} \to \hat{u}_{lpz} = u_{lpz} , \qquad u_{lph} \to \hat{u}_{lph} = u_{lph} . \qquad (29c)$$

Using these transformations (from Eqs.(20a) and (20b)) we replace $\hat{v}(\zeta)$ in Eqs.(15) and (16) (at n=2 and $b_2 \equiv b$) with $(\hat{u}(\zeta) - \hat{u}_{lp}(\zeta) - u_S)/(1-u_S)$, $v_h$ and $v_z$ with $(u_h - u_{lph} - u_S)/(1-u_S)$ and $(u_z - u_{lpz} - u_S)/(1-u_S)$, respectively, and denote $b_u$ to be

$$b_u \equiv (u_h - u_z + u_{lpz} - u_{lph})/\zeta_h^2 . \qquad (30)$$

As a result we obtain the relative volume of the soil intra-aggregate matrix at swelling, $\hat{u}(\zeta)$ as (cf. Eq. (15) for clay swelling at n=2 and $b_2 \equiv b$)

$$\hat{u}(\zeta) = u_h + \hat{u}_{lp}(\zeta) - u_{lph} - b_u(\zeta - \zeta_h)^2 , \qquad 0 \leq \zeta \leq \zeta_h . \qquad (31)$$

The $k(c/c_*)$ lacunar factor does not depend on the water content and direction of its variation (wetting or drying) [14,15]. For this reason Eq.(23) (after accounting for the substitutions, $u_{lp}(\zeta) \to \hat{u}_{lp}(\zeta)$ and $v(\zeta) \to \hat{v}(\zeta)$) and the expressions for $k(c/c_*)$ (Eqs.(24a) and (24b)) are also retained in case of swelling. Replacing $\hat{u}_{lp}(\zeta)$ in Eq.(31) with $\hat{u}_{lp}(\zeta)$ from Eq.(23) (after the substitutions) we obtain the final expression for $\hat{u}(\zeta)$ to be (cf. Eq.(21) for shrinkage of the intra-aggregate matrix)

$$\hat{u}(\zeta) = u_h + u_{lpz} - u_{lph} - k(1-u_S)(\hat{v}(\zeta) - v_z) - b_u(\zeta - \zeta_h)^2, \qquad 0 \leq \zeta \leq \zeta_h. \tag{32}$$

One can check that according to Eq.(32) $\hat{u}(\zeta_h) = u_h$ and $\hat{u}(0) = u_z$, as it should be, in force of Eq.(23) at $\zeta = \zeta_h$ and Eq.(30). The qualitative view of $\hat{u}(\zeta)$ flows out of Eq.(32) and coincides with the view of $\hat{v}(\zeta)$ in Fig.2. The mutual arrangement of the pair, $u(\zeta)$ and $\hat{u}(\zeta)$ also repeats that of $v(\zeta)$ and $\hat{v}(\zeta)$ in Fig.2. However, the initial slope $d\hat{u}/d\zeta_{|\zeta=0}$ differs from $d\hat{v}/d\zeta_{|\zeta=0}$ (Eq.(17); Fig.2). One can obtain $d\hat{u}/d\zeta_{|\zeta=0}$ from Eqs.(32), (16), (17), (30), and Eqs.(21), (23) at $\zeta = \zeta_h$, to be

$$d\hat{u}/d\zeta_{|\zeta=0} = 4(u_h - u_z) = 2(1 + u_s - 2u_z) + 4u_{lph} \tag{33}$$

(cf. Eq.(17)). Thus, $d\hat{u}/d\zeta_{|\zeta=0}$ depends not only on clay type (as $d\hat{v}/d\zeta_{|\zeta=0}$), but also on the clay content, soil texture, and intra-aggregate structure (see the term $4u_{lph}$). To transit to the coordinates $(w, \hat{U})$ (specific volume, $\hat{U}$ at swelling vs. gravimetric water content, $w$ of the soil intra-aggregate matrix), one can use Eq.(27) by replacing $u \to \hat{u}$ and $U \to \hat{U}$. Then the initial slope, $d\hat{U}/dw_{|w=0}$ is given as

$$d\hat{U}/dw_{|w=0} = d\hat{u}/d\zeta_{|\zeta=0}/[(1-u_s)\rho_w] = [2(1+u_s-2u_z) + 4u_{lph}]/[(1-u_s)\rho_w]. \tag{34}$$

$\hat{U}(w)$ (dash line $\hat{1}$ in Fig.3) is qualitatively similar to $\hat{u}(\zeta)$ and $\hat{v}(\zeta)$ (Fig.2).

## 4. The reference swelling curve of an aggregated soil
### 4.1. The major points of an available model in the shrinkage case

(1) The reference shrinkage curve corresponds to the zero contribution of the crack volume and is a single-valued characteristic of soil [13-15]. The reference shrinkage is realized at sufficiently small sample size [17]. In addition to the intra-aggregate matrix, the aggregates of soils include the surface layer (Fig.1) or *interface* layer that is deformable, but non-shrinking. At shrinkage, the intra-aggregate matrix and interface layer determine: (a) two contributions, $w'$ and $\omega$, respectively, to the total water content, $W$ of a soil as

$$W = w' + \omega, \qquad 0 \leq W \leq W_h, \qquad 0 \leq w' \leq w'_h, \qquad 0 \leq \omega \leq \omega_h = W_h - w'_h \tag{35}$$

and (b) two contributions, $U'$ and $U_i$, respectively, to the specific volume of aggregates, $U_a$ as

$$U_a = U' + U_i. \tag{36}$$

The specific volumes of soil, $Y_r$ (index "r" indicates the reference shrinkage), aggregates, $U_a$, and structural pores, $U_s$=const are linked as

$Y_r = U_a + U_s = U' + U_i + U_s$ . (37)

The values, $\omega$, $W$, $U'$, $U_a$, and $Y_r$ depend on the water content contribution, $w'$ of the intra-aggregate matrix to $W$. The $w'$ value is in the range $0 \leq w' \leq w'_h$ (Fig.3, the $w'$ axis) where $w'_h$ corresponds to the state of maximum soil swelling, $W_h$ (Fig.3, the $W$ axis). Thus, to find the reference shrinkage curve, $Y_r(W)$ in the parametric view $(W(w'), Y_r(w'))$ one should know functions, $U'(w')$ (in Eqs.(36) and (37)) and $\omega(w')$ (in Eq.(35)) as well as the constant (for the soil) specific volumes $U_i$ and $U_s$.

(2) The key point for the determination of $U'(w')$ is the simple link between the intra-aggregate matrix contribution to the specific volume of aggregates at shrinkage, $U'(w')$ (dash-dot curve 2 in Fig.3) and the shrinkage curve of the intra-aggregate matrix, $U(w)$ (dash line 1 in Fig.3; for $U(w)$ see the end of Section 3.1) [13,14]. Unlike $U'$ and $w'$ (that relate to unit mass of the oven-dried soil as a whole), $U$ and $w$ are the specific volume and water content of the *same* intra-aggregate matrix per unit mass of the oven-dried matrix itself (but not the soil as a whole; Fig.1). The water contents, $w$ and $w'$ (see the $w$ and $w'$ axes in Fig.3) are connected as $w'=w/K$ and $U$ and $U'$ values (curves 1 and 2 in Fig.3) as $U'=U/K$ where $K$ is the ratio of the aggregate solid mass to the solid mass of the intra-aggregate matrix (i.e., the solid mass of aggregates without the interface layer; see Fig.1). Thus, the auxiliary curve $U'(w')$ (dash-dot curve 2 in Fig.3) is expressed through the shrinkage curve $U(w)$ (dash curve 1 in Fig.3) of the intra-aggregate matrix (Section 3.1) as

$U'(w') = U(w'K)/K$, $\quad 0 \leq w' \leq w'_h \quad (w'=w/K)$ . (38)

(3) For the ways to estimate the $K$ value see [13,14,17]. The most fundamental way to estimate the value of $K$, through the parameters of the soil structure and texture [25], is based on the calculation of $U_i$ and use of the relation between $K$ and $U_i$ as.

$K = (1 - U_i/U_h)^{-1}$ . (39)

Here $U_h$ is the maximum specific volume of the intra-aggregate matrix at $W=W_h$ (Fig.3). The specific volume of the interface layer at shrinkage, $U_i$ is found through the aggregate-size distribution, $F(X, P_h) \equiv F(\eta, P_h)$ at $W=W_h$ ($X$ being the current aggregate size; $P_h$ – inter-aggregate porosity at $W=W_h$; for $\eta$ see below), using a $G(\alpha,\beta,\chi)$ function to be

$$G(\alpha,\beta,\chi) = \chi \left\{ 1 - (1-\alpha)^3 + 3[\alpha/(1-\alpha)] \int_0^1 \frac{\eta^2 F(\eta,\beta) d\eta}{[\eta + \alpha/(1-\alpha)]^4} \right\}$$ (40)

where the $F(\eta, \beta)$ distribution is from [26] as

$F(\eta, \beta) = (1 - \beta^{I_o(\eta)/8.4})/(1-\beta)$, $\quad I_o(\eta) = \ln(6)(4\eta)^4 \exp(-4\eta)$, $\quad 0 \leq \eta \leq 1$ . (41)

In the particular, and important, case of an aggregated soil with negligible structural porosity ($\beta \to 0$) the $F(\eta,\beta)$ distribution in Eqs.(40) and (41) is replaced with

$$F(\eta, 0)=\{1-\exp[-I_o(\eta)]\}/[1-\exp(-8.4)] \ . \tag{42}$$

Taking in Eqs.(40) and (41) ($X_{min}$ and $X_m$ being the minimum and maximum aggregate size at $W=W_h$; $x_n$ being the mean size of soil solids; $x_n \cong X_{min}$)

$$\alpha=x_n/X_m, \quad \beta=P_h, \quad \chi=U_h, \quad \eta=(X-X_{min})/(X_m-X_{min}), \tag{43}$$

one obtains $U_i$ (Fig.3, vertical displacement between curves 2 and 3) as

$$U_i=G(x_n/X_m, P_h, U_h) \ . \tag{44}$$

Thus, $K$ is a function of the $x_n/X_m$ ratio and $P_h$. Figure 3 shows the $U(w)$ (dash curve 1), $U'(w')$ (dash-dot curve 2), and $U_a(w')$ (solid curve 3 at $w'<w'_s$ plus dotted line 3' at $w'>w'_s$) curves at $0 \le w \le w_h$ (the $w$ axis) and $0 \le w' \le w'_h=w_h/K$ (the $w'$ axis).
(4) The key point for determination of the water contribution of the interface layer at shrinkage, $\omega(w')$ is the presentation of $\omega(w')$ [13,14] as

$$\omega(w') = \begin{cases} 0, & 0 \le w' < w'_s \\ \rho_w U_i \Pi_h F_i(\eta(R(w'), R_{min}, R_m), \Pi_h), & w'_s \le w' < w'_h, \end{cases} \tag{45}$$

where the clay porosity of the interface layer, $\Pi_h$ (that coincides with the clay porosity of the intra-aggregate matrix at maximum swelling and is the generalization of Eq.(10) at $\zeta=\zeta_h$) is

$$\Pi_h=1-(u_s+u_{lph})/u_h \ , \tag{46}$$

$w'_s$ corresponds to the end point of structural shrinkage (Fig.3; curve 3; axis $w'$); $F_i(\eta,\Pi_h)$ is the volume fraction of the water-filled interface clay pores at a given $w'$ value at shrinkage. In the simplest case $F_i(\eta,\Pi_h)$ is presented as ($I_o(\eta)$ from Eq.(41))

$$F_i(\eta,\Pi_h)=(1-(1-\Pi_h)^{I_o(\eta)/8.4})/\Pi_h, \qquad \eta(R(w'), R_{min}, R_m)=(R-R_{min})/(R_m-R_{min}). \tag{47}$$

$R_{min}$ and $R_m$ are the minimum and maximum sizes of non-shrinking clay pores in the interface layer matrix (Fig.4; for the $R_{min}$ and $R_m$ values and more complex case of the $F_i(\eta,\Pi_h)$ distribution see [13]); $R=R(w')$ is the maximum size of water-filled clay pores of the intra-aggregate matrix (Fig.4, curve 3; it is convenient to consider $R$ as $R(\zeta)$ at $0 \le \zeta \le \zeta_h$ in Fig.4 and after that to transit to $R(w')$ using Eq.(8) and $w'=c\overline{w}/K$). The known $R_{min}$ value determines the $w'_s$ point from condition, $R(w'_s)=R_{min}$ (Fig.4, point at $\zeta=\zeta_s$). This condition means exhausting water, $\omega(w')$ in the interface clay pores at $w'<w'_s$ (cf. Eq.(45)) and flows out of the capillarity considerations (that at any possible $w'$ the current size of the maximum water-filled clay pores in the interface layer coincides with that in the intra-aggregate matrix at shrinkage, i.e., $R(w')$). The $R(w')$ dependence in the $w'_n \le w' \le w'_h$ range (that contains $w'_s$) is determined as (see curves 1 and 3 in Fig.4)

$$R(w') = r_m(w'K), \qquad w'_n \leq w' \leq w'_h \tag{48}$$

where $r_m(w'K)$ is from Eq.(11a). After finding $\omega(w')$ the $W(w')=w'+\omega(w')$ dependence allows one to transform in Fig.3 the $w'$ axis to $W$ and to correspondingly transform $U_a(w')$ (Fig.3; solid curve 3 at $w'<w'_s$ plus dotted line 3' at $w'>w'_s$) to $U_a(W)$ (Fig.3; solid curve 3). The slope of the reference shrinkage curve, $dU_a/dW=dY_r/dW$ (since $Y_r=U_a+U_s$) in the basic shrinkage range, $W_n \leq W \leq W_s$ (Fig.3) coincides with that in Eq.(28) for the intra-aggregate matrix.

(5) For some relations between $U_s$, $K$, $U_i$, and $W_h$, that can be useful in the calculation of $U'(w')$ and $\omega(w')$, see [14]. In the general case the input physical parameters (in the calculation of the reference shrinkage curve, $Y_r(W)$), reflecting the soil texture, and inter- and intra-aggregate structure, include, $v_s$, $v_z$, $\rho_s$, $c$, $P_z$, $X_m$, $x_n$, $p$, and $Y_{rh}$ [14,17] where $P_z$ is structural porosity in the oven-dried state and $Y_{rh}$ is the specific soil volume at the start of reference shrinkage (at $W=W_h$). All other parameters were explained above (for $p$ see Eq.(25)). Parameters, $Y_{rz}$ and $W_h$ can also be used instead of $v_s$ and $v_z$. In particular cases the number of input parameters is essentially reduced.

### *4.2. Generalization to the swelling case*
### *4.2.1. General presentation of the soil reference swelling curve*

If cracks do not appear and develop in a sufficiently small sample at shrinkage [17] *a fortiori* they do not appear at swelling. Similar to reference shrinkage, such swelling of sufficiently small samples without cracks is referred to below as *reference swelling*. Alternating shrinkage and swelling lead to the gradual destruction of aggregates (see, e.g., [27]). That is, the maximum aggregate size, $X_m$ gradually decreases (while the mean size of soil solids, $x_n$ is retained). The decrease of $X_m$ leads to the formation of the additional aggregate surface, that is, to increasing of the specific interface layer volume and aggregate/intra-aggregate mass ratio [25] compared to their initial values (before shrink-swell cycle), $U_i$ and $K$, respectively. In addition, the maximum water content at swelling, $\hat{W}_h$ cannot be more than $W_h$ at previous shrinkage (see Fig.3). Thus, although in consideration of the reference swelling curve we are based on the same inter- and intra-aggregate structure (Fig.1), the parameter values of the structure *at swelling*, $\hat{X}_m$, $\hat{U}_i$, $\hat{K}$, and $\hat{W}_h$ are, in general, assumed to differ from the values, $X_m$, $U_i$, $K$, and $W_h$ at previous shrinkage as

$$\hat{X}_m \leq X_m, \qquad \hat{U}_i \geq U_i, \qquad \hat{K} \geq K, \qquad \hat{W}_h \leq W_h. \tag{49}$$

Similar to the shrinkage case, accounting for the soil structure (Fig.1) the aggregate volume, $\hat{U}_a$ and total water content, $\hat{W}$ at swelling can obviously be divided into two contributions. In the case of swelling the *volume* and *water* contributions of the intra-aggregate matrix, $\hat{U}'(\hat{w}')$ and $\hat{w}'$, and of interface layer, $\hat{U}_i$ and $\hat{\omega}(\hat{w}')$ give the specific aggregate volume, $\hat{U}_a(\hat{w}')$ to be (cf. Eq.(36))

$$\hat{U}_a(\hat{w}') = \hat{U}'(\hat{w}') + \hat{U}_i, \qquad 0 \leq \hat{w}' \leq \hat{w}'_h \tag{50a}$$

the aggregate water content, $\hat{W}(\hat{w}')$ to be (cf. Eq.(35))

$$\hat{W}(\hat{w}') = \hat{w}' + \hat{\omega}(\hat{w}'), \qquad 0 \leq \hat{w}' \leq \hat{w}'_h \tag{50b}$$

and the soil specific volume, $\hat{Y}_r(\hat{w}')$ at the reference swelling as (cf. Eq.(37))

$$\hat{Y}_r(\hat{w}')=\hat{U}'(\hat{w}')+\hat{U}_i+U_s \qquad (51)$$

(the specific volume of structural pores, $U_s$ at the reference shrinkage and swelling is the same and constant). Now, similar to Eq.(38) (the shrinkage case), $\hat{U}'(\hat{w}')$ is

$$\hat{U}'(\hat{w}')=\hat{U}(\hat{w}'\hat{K})/\hat{K}, \quad 0 \leq \hat{w}' \leq \hat{w}'_h \qquad (\hat{w}'=w/\hat{K}) \qquad (52)$$

where $\hat{U}$ is the specific volume of the intra-aggregate matrix at swelling (see the end of section 3.2). The water content, $w$ of the intra-aggregate matrix (per unit solid mass of the matrix itself) is the same at shrinkage and swelling, i.e., $w \equiv \hat{w}$ (Fig.3; $w$ axis) (similar to clay shrinkage-swelling, where $\overline{w} \equiv \hat{\overline{w}}$ and $\zeta \equiv \hat{\zeta}$, as in Fig.2).

In addition to curves $U(w)$ (curve 1), $U'(w')$ (curve 2), and $U_a(w')$ (in part curve 3 and in part line 3'), relating to shrinkage and *convex downward*, Fig.3 qualitatively shows the similar curves, $\hat{U}(w)$ (curve $\hat{1}$), $\hat{U}'(\hat{w}')$ (curve $\hat{2}$), and $\hat{U}_a(\hat{w}')$ (dotted curve $\hat{3}'$ and solid curve $\hat{3}$ at $\hat{w}'<\hat{w}'_h$) relating to swelling and *convex upward*. $\hat{U}_a(\hat{w}')$ (Eq.(50a)) (or $\hat{Y}_r(\hat{w}')$, Eq.(51)) and $\hat{W}(\hat{w}')$ (Eq.(50b)) together give a parametric presentation of $\hat{U}_a(\hat{W})$ (Fig.3; solid curve $\hat{3}$) (or $\hat{Y}_r(\hat{W})$). Figure 3 does not show the reference swelling curve, $\hat{Y}_r=\hat{U}_a+U_s$ because at $U_s$=const $\hat{Y}_r(\hat{W})$ is just parallel to the $\hat{U}_a(\hat{W})$ curve. Thus, similar to the shrinkage case one should find $\hat{U}'(\hat{w}')$, $\hat{U}_i$, and $\hat{\omega}(\hat{w}')$ entering Eqs.(50a) and (50b). In Sections 4.2.2-4.2.4 we show that $\hat{U}'(\hat{w}')$, $\hat{U}_i$, and $\hat{\omega}(\hat{w}')$, and, hence, the soil reference swelling curve, are determined in a single-valued manner by the characteristics of previous shrinkage and aggregate sizes after their destruction. First, we consider the volume contributions, $\hat{U}'(\hat{w}')$ and $\hat{U}_i$ to soil reference swelling (Sections 4.2.2-4.2.3).

*4.2.2. Relation of the mass ratio ($\hat{K}$) and interface layer volume ($\hat{U}_i$) at swelling*

According to Eq.(52) we need to know the aggregate/intra-aggregate mass ratio at swelling, $\hat{K}$. To find $\hat{K}$ one should first derive the relation between $\hat{K}$ and the interface layer volume at swelling, $\hat{U}_i$, that is, the generalization of Eq.(39) relating to shrinkage. The relation being sought should depend on the characteristics of previous shrinkage, such as $U_z$, $U_h$, (see Fig.3) and $K$ (or $U_i$). Replacing $U'$ in Eq.(36) with $U/K$ one can obtain at $w'$=0 (marked by index "z") $U_{az}=U_z/K+U_i$. Similarly, replacing in Eq.(50a) $\hat{U}'$ with $\hat{U}/\hat{K}$, one can obtain at $\hat{w}'$=0 (marked by index "z") $\hat{U}_{az}=\hat{U}_z/\hat{K}+\hat{U}_i$. That can be rewritten as $U_{az}=U_z/\hat{K}+\hat{U}_i$ since $\hat{U}_{az}=U_{az}$ and $\hat{U}_z=U_z$ (see Fig.3). The difference between $U_{az}=U_z/\hat{K}+\hat{U}_i$ and $U_{az}=U_z/K+U_i$ gives the additional interface layer volume, $\Delta\hat{U}_i \equiv \hat{U}_i - U_i$ that appears at swelling as

$$\Delta\hat{U}_i \equiv \hat{U}_i - U_i = U_z(1/K - 1/\hat{K}) . \qquad (53)$$

It follows that the relation sought for $\hat{K}$ is ($\hat{K} \to K$ at $\hat{U}_i \to U_i$ as it should be)

$$\hat{K}=K[1-K(\hat{U}_i-U_i)/U_z]^{-1} \quad . \tag{54}$$

Thus, to find $\hat{K}$ according to Eq.(54), besides $K$, $U_z$, $U_h$, and $U_i$ (that are known after the shrinkage stage [13,14]), we should preliminarily calculate $\hat{U}_i$. Similar to $U_i$ at shrinkage [25], $\hat{U}_i$ at swelling can be connected with a corresponding aggregate-size distribution that originates from the aggregate destruction at swelling and is characterized by $\hat{X}_m \leq X_m$ (Eq.(49)).

*4.2.3. The interface layer volume ($\hat{U}_i$) and aggregate structure at swelling*

Estimation of $U_i$ is reached by Eqs.(40)-(44). In a similar way one can also estimate $\hat{U}_i$ by only replacing the arguments, α, β, χ, η in Eq.(43) with

$$\alpha=x_n/\hat{X}_{mz} , \quad \beta=P_z , \quad \chi=U_z , \quad \eta=(X-X_{min})/(\hat{X}_{mz}-X_{min}) \tag{55}$$

where $\hat{X}_{mz}$ is the maximum aggregate size after destruction at swelling close to $\hat{W}=0$; $P_z=\hat{P}_z$ is the structural porosity at swelling close to $\hat{W}=0$; and $U_z=\hat{U}_z$ is the specific volume of the intra-aggregate matrix at swelling close to $\hat{W}=0$. Indeed, the specific volume, $\hat{U}_i$ of the interface layer at swelling (see Fig.3) includes the previous specific volume, $U_i$ (that we know) and additional interface layer volume, $\Delta\hat{U}_i=\hat{U}_i - U_i$ that appeared as a result of aggregate destruction (see section 4.2.1) and the formation of additional aggregate surfaces. To estimate $\hat{U}_i$ using the $G$ function from Eq.(40), one should specify some features of the aggregate destruction. We *assume* that the latter occurs during a short *initial* stage of swelling under the action of increasing pore pressure inside aggregates at the expense of air entrapped and compressed in the pores of contributive clay at soil wetting. That is, the aggregate destruction and formation of the additional aggregate surfaces and interface layer volume, $\Delta\hat{U}_i$ at reference swelling occurs in the *beginning* of swelling, in the state close to $\hat{W}=0$ with corresponding $\hat{U}=\hat{U}_z=U_z$ (Fig.3, Eq.(55)), $\hat{P}=\hat{P}_z=P_z$, and the maximum aggregate size after destruction, $\hat{X}_{mz}$. It is worth remembering that the interface layer volume before aggregate destruction, $U_i$ (Fig.3) which is retained after destruction and contributes to $\hat{U}_i$, was formed in the state of maximum swelling at the specific intra-aggregate volume $U=U_h$ (Fig.3, Eq.(43)), $P=P_h$, and $W=W_h$ [13,14]. That is, the pore structure of the $U_i$ and $\Delta\hat{U}_i$ contributions to $\hat{U}_i$ is *different* (see Section 4.2.4). However, the mean interface layer thickness, $x_n/2$ [25] is the *same* for the existing, $U_i$ and additional, $\Delta\hat{U}_i$ contributions to $\hat{U}_i$. Accounting for this fact one can find $\hat{U}_i$ by a totally similar method as that in the case of $U_i$ [25] (with replacements from Eq.(55)). Thus, we come to the calculation of $\hat{U}_i$ by Eqs.(40)-(42), (55), and replacement of Eq.(44) with

$$\hat{U}_i=G(x_n/\hat{X}_{mz}, P_z, U_z) \quad . \tag{56}$$

Parameters $x_n$, $P_z$, and $U_z$ in this link between $\hat{U}_i$ and $\hat{X}_{mz}$ after aggregate destruction, are known from the shrinkage stage. The relation should be noted between the maximum aggregate size at swelling in its beginning ($\hat{W}=0$), $\hat{X}_{mz}$ entering Eq.(56) and at the end of swelling ($\hat{W}=\hat{W}_h$), $\hat{X}_m$. The $\hat{X}_{mz}$ and $\hat{X}_m$ sizes are interconnected as

$$\hat{X}_m = x_n + (\hat{X}_{mz} - x_n)(u_h/u_z)^{1/3} \qquad (57)$$

where $u_z$ and $u_h$ are the relative volume of the soil intra-aggregate matrix at the shrinkage limit and maximum swelling, respectively, that are found in the course of the shrinkage curve prediction [13, 14] ($u_h/u_z = U_h/U_z$). The relation between $\hat{X}_{mz}$ and $\hat{X}_m$ at swelling (Eq.(57)) and its derivation are totally similar to the relation between $X_{mz}$ and $X_m$ at shrinkage and its derivation [25]. The physical link between the maximum aggregate sizes before ($X_{mz}$ or $X_m$) and after ($\hat{X}_{mz}$ or $\hat{X}_m$) destruction at swelling, should include, in addition to the characteristics of previous shrinkage (see above) and swelling conditions (see paragraph after Eq.(15)) also the strength characteristics of the aggregates before destruction. For this reason $\hat{X}_{mz}$ enters the model as an independently measured characteristic (like $X_{mz}$). Estimating $\hat{X}_{mz}$ see in Section 6.4.

Using the concepts of the interface layer volume, $U_i$ and $\hat{U}_i$ one can estimate the possible *variation* of the *maximum swelling volume*, $U_{ah} - \hat{U}_{ah}$ after the *reference* shrink-swell cycle (see Fig.3; $U_{ah} = U_a(w'_h)$ and $\hat{U}_{ah} = \hat{U}_a(\hat{w}'_h)$). By definition of $U_a$ (Fig.3) and $\hat{U}_a$ (Eq.(50a)), $U_{ah} = U_h$ and $\hat{U}_{ah} = \hat{U}_h/\hat{K} + \hat{U}_i = U_h/\hat{K} + \hat{U}_i$ ($\hat{U}_h = U_h$; see Fig.3). Substituting for $U_{ah}$ and $\hat{U}_{ah}$ values, we have

$$U_{ah} - \hat{U}_{ah} = U_h - \hat{U}_i - U_h/\hat{K} . \qquad (58a)$$

Replacing $\hat{K}$ with its expression from Eq.(54), then replacing $K$ with its expression from Eq.(39) and after that, rearranging the right part of Eq.(58a) we obtain

$$U_{ah} - \hat{U}_{ah} = (U_h/U_z - 1)(\hat{U}_i - U_i) . \qquad (58b)$$

Thus, according to the model the *maximum-swelling-volume decrease* after the *reference* shrink-swell cycle, $U_{ah} - \hat{U}_{ah}$ (Eq.(58b)) is totally determined: (i) by the existence of the interface layer (i.e., $U_i > 0$ at shrinkage and $\hat{U}_i > 0$ at swelling) and (ii) by the increase of the interface layer volume at swelling (i.e., $\hat{U}_i > U_i$). Hence, the observation itself of maximum-swelling-volume decrease after the soil shrink-swell cycle, $U_{ah} - \hat{U}_{ah} > 0$ (e.g. [1]) is the strong qualitative experimental evidence in favor of both the existence of the interface layer with specific properties and the increase of interface layer volume at swelling (compared to that at previous shrinkage). Note that $\hat{X}_{\text{mz}}$ can be estimated from the experimental data on the decrease in maximum swelling volume (Eqs.(58b) and (56)).

*4.2.4. Relations between water contributions to soil reference swelling*

Now we can proceed to estimating the water contribution, $\hat{\omega}(\hat{w}')$ of the interface layer after aggregate destruction, to the total water content at swelling, $\hat{W}$ (see Eq.(50b)). After aggregate destruction (i.e., at swelling) the contribution of the interface layer volume, $\hat{U}_i$ to $\hat{U}_a(\hat{w}')$ (Eq.(50a)) is divided into the interface layer volume contribution, $U_i$ before the aggregate destruction and additional volume contribution, $\Delta \hat{U}_i$ of the interface layer after the aggregate destruction. Similarly, the interface layer *water* contribution, $\hat{\omega}$ to $\hat{W}$ at swelling (Eq.(50b)) can be presented as

$$\hat{\omega} = \omega + \Delta\hat{\omega} \quad , \tag{59}$$

with $\omega$ being the water contribution of the interface layer that existed before aggregate destruction (with the specific volume, $U_i$) and $\Delta\hat{\omega}$ being the water contribution of the additional interface layer that came into being after aggregate destruction (with the specific volume, $\Delta\hat{U}_i$). However, unlike $U_i$ and $\Delta\hat{U}_i$, $\omega$ and $\Delta\hat{\omega}$ depend on the water content contribution, $\hat{w}'$ of the intra-aggregate matrix to $\hat{W}$ at swelling (i.e., after aggregate destruction). In addition, unlike $U_i$, $\omega$ as a function of water content, $w$ of intra-aggregate matrix (Eq.(27)) at shrinkage (Eqs.(45)-(48)) and swelling (see below) is different. Before estimation of $\omega$ and $\Delta\hat{\omega}$ it is worth noting the simple relations between different water axes (the $w$, $w'$, $\hat{w}'$, $W$, and $\hat{W}$ axes are shown in Fig.3). For the relation: relative water content ($\zeta$) – gravimetric water content of clay ($\overline{w}$), see Eq.(8). The gravimetric water content of an intra-aggregate matrix ($w$) and that of a contributive clay ($\overline{w}$) relate as $w = c\overline{w}$. For the relation: relative water content ($\zeta$) – gravimetric water content of intra-aggregate matrix ($w$), see Eq.(27). The gravimetric water content of the intra-aggregate matrix at shrinkage ($w$) and swelling ($\hat{w}$) coincide as (hence $\zeta = \hat{\zeta}$ and $\overline{w} = \hat{\overline{w}}$ also)

$$w = \hat{w} \quad . \tag{60}$$

The water contents at shrinkage, $w'$ and $w$, and at swelling, $\hat{w}'$ and $\hat{w}$ are linked as

$$w = w'K \quad \text{and} \quad \hat{w} = \hat{w}'\hat{K} \quad . \tag{61}$$

Since $K/\hat{K} < 1$ (see Eq.(49)), according to Eqs.(60) and (61)

$$\hat{w}' = (K/\hat{K})w' < w' \quad . \tag{62}$$

To estimate $\omega(\hat{w}')$ at swelling we can use Eq.(45) with replacements, $w' \to \hat{w}'$, $w'_s \to \hat{w}'_b$ (see $w'_s$ on the $w'$ axis and $\hat{w}'_b$ on the $\hat{w}'$ axis in Fig.3; $\hat{w}'_b$ corresponds to the *beginning* point of water filling in the interface layer part of the $U_i$ volume at swelling; $\zeta_b$ in Fig.4 corresponds to $\hat{w}'_b$), and $w'_h \to \hat{w}'_h$ (see $w'_h$ on the $w'$ axis and $\hat{w}'_h$ on the $\hat{w}'$ axis in Fig.3 and corresponding $\zeta_h$ in Fig4; $\hat{w}'_h$ is connected with $w'_h$ by Eq.(62)). In addition, the maximum internal size of the water-filled clay pores of intra-aggregate matrix at shrinkage, $R(w')$ from Eq.(48) (see Fig.4, curve 3) is replaced with another dependence at swelling, $\hat{R}(\hat{w}')$ (see Fig.4, curve 6) that will be considered below (Fig.4 shows $\hat{R}$ as a function of $\zeta$ for convenience; in order to transit to $\hat{w}'$ one should use Eq.(8), $w'=c\overline{w}/K$, and Eq.(62)). As a result we have

$$\omega(\hat{w}') = \begin{cases} 0, & 0 \le \hat{w}' < \hat{w}'_b \\ \rho_w U_i \Pi_h F_i(\eta(\hat{R}(\hat{w}'), R_{\min}, R_m), \Pi_h), & \hat{w}'_b \le \hat{w}' < \hat{w}'_h, \end{cases} \quad (63)$$

with the same $U_i$ (Eq.(44)), $\Pi_h$ (Eq.(46)), $F_i(\eta,\Pi_h)$ (Eq.(47) in the simplest case), $I_o(\eta)$ (Eq.(41)), $R_{\min}$, $R_m$ again from [13] (see also Fig.4), and

$$\eta(\hat{R}(\hat{w}'), R_{\min}, R_m) = (\hat{R}(\hat{w}') - R_{\min})/(R_m - R_{\min}) \quad (64)$$

(cf. Eq.(47)). With that, $\omega(\hat{w}')$ varies in the range (see $\omega_h$ in Fig.3)

$$0 \le \omega(\hat{w}') \le \omega_h = \omega(\hat{w}'_h) = \rho_w U_i \Pi_h, \qquad \hat{w}'_b \le \hat{w}' \le \hat{w}'_h \ . \quad (65)$$

Finally, it is necessary to note that the $\hat{w}'_b$ point is determined from the condition, $\hat{R}(\hat{w}'_b) = R_{\min}$ (Fig.4, point at $\zeta=\zeta_b$). This condition means the beginning of water-filling in the clay pores (of the interface layer part of the $U_i$ volume) at $\hat{w}' > \hat{w}'_b$ and flows out of the capillarity considerations, that at any possible $\hat{w}'$ the current size of the maximum water-filled clay pores in the interface layer coincides with that in the intra-aggregate matrix at swelling, i.e., $\hat{R}(\hat{w}')$ (cf. this determination of $\hat{w}'_b$ with that of $w'_s$ in the text being between Eqs.(47) and (48)).

To estimate $\Delta\hat{\omega}(\hat{w}')$ at swelling (entering Eq.(59)) we can use the following relations (cf. Eqs.(45)-(47) and Eqs.(63)-(65))

$$\Delta\hat{\omega}(\hat{w}') = \begin{cases} \rho_w \Delta\hat{U}_i \Pi_z F_i(\eta(\hat{R}(\hat{w}'), \hat{R}_{\min}, \hat{R}_m), \Pi_z), & 0 \le \hat{w}' \le \hat{w}'_e \\ \rho_w \Delta\hat{U}_i \Pi_z, & \hat{w}'_e \le \hat{w}' \le \hat{w}'_h \end{cases} \quad (66)$$

where (cf. Eq.(46))

$$\Pi_z = 1 - (u_s + u_{lpz})/u_z \ , \quad (67)$$

the $F_i(\eta,\Pi_z)$ function coincides with Eq.(47) at the same $I_o(\eta)$ from Eq.(41) (but with another $\eta$; see below) and after replacement $\Pi_h \to \Pi_z$; $\hat{R}(\hat{w}')$ is the maximum internal size of the water-filled clay pores of the intra-aggregate matrix at swelling (see Fig.4, curve 6) that is considered below; $\hat{R}_{\min}$ and $\hat{R}_m$ are the minimum and maximum sizes of non-shrinking and non-swelling clay pores in the interface layer part of the $\Delta\hat{U}_i$ volume at swelling (Fig.4); $\hat{R}_{\min}=r_o(\zeta=0)$ (see Eq.(11b); Fig.4); $\hat{R}_m=r_m(\zeta=0)$ (see Eq.(11a); Fig.4); $\eta(\hat{R}(\hat{w}'),\hat{R}_{\min},\hat{R}_m)$ follows Eq.(64) after replacements, $R_{\min}\rightarrow\hat{R}_{\min}$ and $R_m\rightarrow\hat{R}_m$; and $\hat{w}'_e$ corresponds to the *end* point of water filling in the interface layer part of the $\Delta\hat{U}_i$ volume at swelling (see $\hat{w}'_e$ on axis $\hat{w}'$ in Fig.3; in Fig.4 $\hat{w}'_e\rightarrow\zeta_e$). According to Eq.(66) $\Delta\hat{\omega}(\hat{w}')$ varies in the range (see $\Delta\hat{\omega}_e$ in Fig.3)

$$0\leq\Delta\hat{\omega}(\hat{w}')\leq\Delta\hat{\omega}_e=\Delta\hat{\omega}(\hat{w}'_e)=\rho_w\Delta\hat{U}_i\Pi_z, \qquad 0\leq\hat{w}'\leq\hat{w}'_e. \qquad (68)$$

The $\hat{w}'_e$ point is determined from the condition, $\hat{R}(\hat{w}'_e)=\hat{R}_m$ (Fig.4, point at $\zeta=\zeta_e$). This condition means the end of water-filling in the clay pores (of the interface layer part of the $\Delta\hat{U}_i$ volume) at $\hat{w}'>\hat{w}'_e$ and flows out of the same capillarity considerations as noted above for determining $w'_s$ and $\hat{w}'_b$

Now, we should consider finding the $\hat{R}(\hat{w}')$ function (Fig.4, curve 6). It is convenient to consider $\hat{R}$ as $\hat{R}(\zeta)$. Then, one can transform the water content variable, $\zeta\rightarrow w\rightarrow\hat{w}'$ (see above). It is natural to assume that air, which is entrapped and compressed at clay wetting and swelling in the intra-aggregate matrix, occupies clay pores up to some maximum size (depending on water content), and absorbed water occupies larger pores up to the maximum size, $\hat{R}(\zeta)$ that we are interested in. Then, one can write two different expressions for the summary volume fraction of the entrapped air and absorbed water of the total air-water volume (or of the total pore volume) in the clay. Note that, in general, the total air-water volume also includes the air volume in pores that are larger in size than $\hat{R}(\zeta)$ and under atmospheric pressure. These two different expressions depend on the water content, $\zeta$ of the clay contributing to the soil intra-aggregate matrix. Their equalizing gives the following equation relative to $\hat{R}$ as a function of $\zeta$ (Fig.4, curve 6)

$$f(\eta(\hat{R}(\zeta)),\hat{P})=(1-v_s)\zeta/(\hat{v}(\zeta)-v_s)+(\hat{v}(\zeta)-v(\zeta))/(\hat{v}(\zeta)-v_s), \qquad 0\leq\zeta\leq\zeta_h. \qquad (69)$$

The $f(\eta(\hat{R}(\zeta)),\hat{P})$ function in Eq.(69) is the volume fraction of clay pores of maximum size, $\hat{R}(\zeta)$. Relying on the intersecting-surfaces approach [26] $f(\eta,\hat{P})$ can be written as (cf. Eq.(47))

$$f(\eta,\hat{P})=(1-(1-\hat{P})^{I_o(\eta)/8.4})/\hat{P} \qquad (70)$$

where $I_o(\eta)$ from Eq.(41); $\hat{P}$ is clay porosity at swelling (cf. Eq.(10))

$$\hat{P}(\zeta)=1-v_s/\hat{v}(\zeta), \qquad 0\leq\zeta\leq\zeta_h \qquad (71)$$

and $\eta(\hat{R}(\zeta))$ to be

$$\eta(\hat{R}(\zeta))=(\hat{R}(\zeta)-\hat{r}_o(\zeta))/(\hat{r}_m(\zeta)-\hat{r}_o(\zeta)). \qquad (72)$$

Here $\hat{r}_o(\zeta)$ and $\hat{r}_m(\zeta)$ are the minimum and maximum clay pore sizes at swelling (see Fig.4; curves 5 and 4, respectively). The $\hat{r}_m(\zeta)$ and $\hat{r}_o(\zeta)$ dependences repeat Eqs.(11a) and (11b) with replacement $v(\zeta) \to \hat{v}(\zeta)$ (see Fig.2).

The two terms in the right part of Eq.(69) give the summary volume fraction of absorbed water and entrapped air. The first term relates to water volume fraction according to Eq.(18). This equation (Eq.(18)) is general and not connected with a particular clay pore structure [19]. Note that, unlike $f(\eta, \hat{P})$ in Eq.(69), the saturation degree at swelling, $\hat{F}(\zeta)$, in general, also includes the contribution of water film in pores with size larger than $\hat{R}(\zeta)$. We neglect the volume of the water film in such large pores compared to water volume in water-filled pores with size smaller than $\hat{R}(\zeta)$. This approximation in Eq.(69) for $\hat{R}(\zeta)$ is quite good since we are interested in the $\hat{R}(\zeta)$ solution at sufficiently large $\zeta \sim \zeta_e$ and $\zeta \sim \zeta_b$ (see Fig.4). The second term in the right part of Eq.(69) gives the entrapped-air volume fraction, since $\hat{v} - v_s$ is proportional to the total clay pore volume (at a given $\zeta$), and $\hat{v} - v$ is proportional to the volume of entrapped air. Indeed, at any possible $\zeta$ the volume of solids in $\hat{v}$ and $v$ (Fig.2) is the same, and the volume of the air under atmospheric pressure can be in good approximation considered to be similar at swelling ($\hat{v}$) and shrinkage ($v$).

Thus, $\hat{R}(\zeta)$ (Fig.4, curve 6) is found as a (numerical) solution of Eq.(69) accounting for Eqs.(70)-(72) and the above remarks. After finding $\hat{R}(\zeta)$ one calculates $\omega(\hat{w}')$ (Eq.(63) and accompanying equations), $\Delta\hat{\omega}(\hat{w}')$ (Eq.(66) and accompanying equations), and then $\hat{W}(\hat{w}') = \hat{w}' + \omega(\hat{w}') + \Delta\hat{\omega}(\hat{w}')$ (Eq.(50b)). Finally, $\hat{U}_a(\hat{w}')$ (Section 4.2.2) and $\hat{W}(\hat{w}')$ give the parametric presentation of the reference swelling curve (solid curve $\hat{3}$ in Fig.3). As one can be convinced from the above, the prediction of the reference swelling curve requires the same input physical parameters as those for the prediction of the reference shrinkage curve (solid curve 3 in Fig.3; see the end of Section 4.1). In particular, the initial slope of the reference swelling curve for aggregated soil (Fig.3), $d\hat{U}_a / d\hat{W}|_{\hat{W}=0} = d\hat{Y}_r / d\hat{W}|_{\hat{W}=0}$ (since $\hat{Y}_r = \hat{U}_a + U_s$) depends on all the input parameters and is numerically found because it cannot be presented in a simple form (like Eq.(34) for the intra-aggregate matrix).

## 5. The swelling curve of an aggregated soil with cracks

The model of the cracked-soil swelling curve, $\hat{Y}(\hat{W})$, to be developed, is constructed as a generalization of the recent model of the aggregated-soil shrinkage curve with crack contribution, $Y(W)$ [17]. The interrelations between the sought $\hat{Y}(\hat{W})$ dependence, corresponding shrinkage curve, $Y(W)$, and reference shrinkage ($Y_r(W)$) and swelling ($\hat{Y}_r(\hat{W})$) curves (Section 4) as well as the crack volume contributions are schematically illustrated in Fig.5.

### 5.1. *The major points of an available model in the shrinkage case*

(1) The transition from the reference shrinkage curve of an aggregated soil, $Y_r(W)$ to the shrinkage curve with crack contribution, $Y(W)$ for sufficiently large samples (Fig.5) [17] influences the presentation of the soil volume, but not that of water content. The latter (for $W$) again includes the contributions of the intra-aggregate matrix, $w'$ and interface layer, $\omega(w')$ (Eq.(35); Fig.1). The basic relation for the specific soil volume at shrinkage with cracking, $Y$ includes contributions of the intra-aggregate matrix, $U'$, interface layer, $U_i$, and cracks, $U_{cr}$ as

$$Y(w')=U'(w')+U_i+U_{cr}(w') \ . \qquad 0\leq w'\leq w'_h \ . \qquad (73)$$

The cracks develop from the inter-aggregate (structural) pores with volume, $U_s$ at $w'=w'_h$ (or $W=W_h$ in Fig.5) and $U_{cr}(w')$ meets the initial condition as

$$U_{cr}(w'_h)=U_s \ . \qquad (74)$$

In fact, the crack volume increment is $U_{cr}-U_s$. Equation (73) generalizes the presentation, $Y_r=U'+U_i+U_s$ for the reference shrinkage (i.e., without cracks).
(2) The differential form of Eq.(73) as

$$dY(w')=dU'(w')+dU_{cr}(w') \ , \qquad 0\leq w'\leq w'_h \qquad (75)$$

enables the introduction and definition of the *crack factor*, $q$ [17] using relations

$$dU_{cr}(w')=-qdU'(w') \quad \text{and} \quad dY(w')=(1-q)dU'(w') \ . \qquad (76)$$

(3) The crack factor, $q$ exists in the sample case, $q_s$ and layer one, $q_l$, and, in both variants, depends on the *initial* sample size or layer thickness, $h$, and the characteristics of the aggregate-size distribution at $W=W_h$ (the maximum and minimum aggregate sizes, and inter-aggregate porosity) [17] as

$$q_s(h/h^*)=0, \qquad 0<h/h^*\leq 1 \qquad (77a)$$

$$q_s(h/h^*)=b_1(h/h^*-1)^2, \qquad 1\leq h/h^*\leq 1+\delta \qquad (77b)$$

$$q_s(h/h^*)=1-b_2/(h/h^*-1), \qquad h/h^*\geq 1+\delta \qquad (77c)$$

and

$$q_l(h/h^*)=b_1(h/h^*)^2, \qquad 0\leq h/h^*\leq \delta \qquad (78a)$$

$$q_l(h/h^*)=1-b_2/(h/h^*), \qquad h/h^*\geq \delta \ . \qquad (78b)$$

The theoretical estimates of the universal constants, $b_1\cong 0.15$, $b_2\cong 1$, and $\delta\cong 1.5$, were validated based on the data from [7]. The critical sample size or layer thickness, $h^*\cong 2$-5cm is determined as [17] (where $l_{min}$ and $l_m$ are the mean distances between the aggregates of the minimum, $X_{min}$ and maximum, $X_m$ size, respectively, at $W=W_h$)

$$h^*=10^3(X_m/h^*_o)^3 h^*_o , \qquad h^*_o=(l_{min}l_m)^{1/2} \ . \qquad (79)$$

(4) The integration of the first Eq.(76), using Eq.(74) and $U'(w')=U(w)/K$ (Eq.(38); $w=w'K$), gives the presentation of the crack volume, $U_{cr}(w')$ through $U(w)$ as

$$U_{cr}(w')=q\ (U_h-U(w))/K+U_s , \qquad 0\leq w'\leq w'_h, \quad 0\leq w\leq w_h \ . \qquad (80)$$

Then Eqs.(73) and (80) lead to the presentation of $Y(w')$ through $U(w)$ as

$$Y(w')=(1-q)U(w)/K+qU_h/K+U_s+U_i, \qquad 0\leq w'\leq w'_h, \quad 0\leq w\leq w_h. \tag{81}$$

In the sample case in Eqs.(80) and (81) $q\equiv q_s$, $U_{cr}\equiv U_{cr\,s}$, $Y\equiv Y_s$, and in the layer case $q\equiv q_l$, $U_{cr}\equiv U_{cr\,l}$, $Y\equiv Y_l$. Equation (80) gives at $W=w'=0$ (i.e., at the end of shrinkage), the specific crack volume, $U_{crz}\equiv U_{cr}(0)\equiv Y_z-Y_{rz}+U_s$ as (see Fig.5)

$$U_{crz}=q\,(U_h-U_z)/K+U_s. \tag{82}$$

Accounting for the slope of $U(w)$ in the basic shrinkage range (Eq.(28)) [14], Eq.(81) gives the slope, $S$ of the shrinkage curve $Y(W)$ with crack contribution in the similar range [17] as ($k$ is the lacunar factor [24]; $\rho_w$ is the water density)

$$S=(1-q)(1-k)/\rho_w, \qquad W_n\leq W\leq W_s. \tag{83}$$

(5) Replacing in Eqs.(80) and (81) $U/K$ from $Y_r=U/K+U_i+U_s$ (Eqs.(37) and (38)) one can express the specific crack volume at shrinkage, $U_{cr}(W)$ and the shrinkage curve with cracks, $Y(W)$ through the reference shrinkage curve, $Y_r(W)$ and the $q$ factor as

$$U_{cr}(W)=-qY_r(W)+qY_{rh}+U_s, \qquad 0\leq W\leq W_h \tag{84}$$

$$Y(W)=Y_r(W)+U_{cr}(W)-U_s=(1-q)Y_r(W)+qY_{rh}, \qquad 0\leq W\leq W_h. \tag{85}$$

In Eqs.(84) and (85) $q\equiv q_s$, $U_{cr}\equiv U_{cr\,s}$, and $Y\equiv Y_s$ or $q\equiv q_l$, $U_{cr}\equiv U_{cr\,l}$, and $Y\equiv Y_l$.

### 5.2. Generalization to the swelling case

Similar to transition, $Y_r(W)\rightarrow Y(W)$ for the soil shrinkage curve (Fig.5) (see section 5.1) the transition from the soil reference swelling curve, $\hat{Y}_r(\hat{W})$ (Fig.5) to the soil swelling curve with cracks, $\hat{Y}(\hat{W})$ (Fig.5) that we are interested in, keeps the relations between the water contributions to the total soil water content, $\hat{W}$ at reference swelling (see Sections 4.2.1 and 4.2.4), in particular, the relation given by Eq.(50b), and modifies the major relation between the volume contributions to $\hat{Y}(\hat{W})$ as

$$\hat{Y}(\hat{w}')=\hat{U}_a(\hat{w}')+\hat{U}_{cr}(\hat{w}')=\hat{U}'(\hat{w}')+\hat{U}_i+\hat{U}_{cr}(\hat{w}'), \qquad 0\leq\hat{w}'\leq\hat{w}'_h \tag{86}$$

($\hat{U}'(\hat{w}')$, $\hat{U}_i$, and $\hat{U}_{cr}(\hat{w}')$ being the contributions of the intra-aggregate matrix, interface layer, and crack volume, respectively, at swelling) compared to the similar major relation for $\hat{Y}_r(\hat{w}')$ (Eq.51). $\hat{U}_i$ and $\hat{U}'(\hat{w}')$ in Eqs.(86) and (51) are known from Sections 4.2.1-4.2.3 (Eqs.(56), (52), (27) with replacement $u\rightarrow\hat{u}$ and $U\rightarrow\hat{U}$, Eq.(54)). $\hat{U}_{cr}(\hat{w}')$ and $\hat{Y}(\hat{w}')$ entering Eq.(86) should meet the obvious boundary conditions (Fig.5). Since swelling with cracks ($\hat{Y}(\hat{w}')$) starts after finishing shrinkage with cracks ($Y(w')$) at $\hat{w}'=w'=0$, the soil volumes, $\hat{Y}$ and $Y$ as well as crack volumes, $\hat{U}_{cr}$ and $U_{cr}$ should coincide at this point (Fig.5). These two conditions are not independent because $Y_r$ and $\hat{Y}_r$ also coincide at $\hat{w}'=w'=0$ (Fig.5). Below we use the condition for $\hat{U}_{cr}$ as

$$\hat{U}_{cr}(\hat{w}'=0)=U_{cr}(w'=0)\equiv U_{crz} \tag{87}$$

where the $U_{crz}$ value is known from Eq.(82). To find $\hat{U}_{cr}(\hat{w}')$ and then $\hat{Y}(\hat{w}')$ we will rewrite Eq.(86) in a differential form (cf. Eq.(75)) as

$$d\hat{Y}(\hat{w}')=d\hat{U}'(\hat{w}')+d\hat{U}_{cr}(\hat{w}'), \qquad 0\leq \hat{w}'\leq \hat{w}'_h \tag{88}$$

and define the *crack factor at swelling*, $\hat{q}$ using the relations similar to Eq.(76) for $d\hat{U}_{cr}(\hat{w}')$, $d\hat{U}'(\hat{w}')$, and $d\hat{Y}(\hat{w}')$. By *definition* $\hat{q}$ again is the fraction of the increment of the aggregate volume at swelling, $d\hat{U}_a(\hat{w}')=d\hat{U}'(\hat{w}')>0$ (see Eq.(86) where $\hat{U}_i$=const) that is transformed to the corresponding increment of the crack volume inside the soil, $d\hat{U}_{cr}(\hat{w}')<0$. However, one should take into account that the crack factor does not depend on water content and, in particular, on the direction of its variation, that is, wetting or drying (cf. a similar statement as applied to the lacunar factor, $k$ in Section 3.2). For this reason, for a given soil and initial sample size (before shrink-swell cycle) the crack factor at shrinkage and swelling stages of the shrink-swell cycle should coincide, $\hat{q}=q$. Thus, the relations that are similar to Eq.(76) for the swelling stage are

$$d\hat{U}_{cr}(\hat{w}')=-qd\hat{U}'(\hat{w}') \quad \text{and} \quad d\hat{Y}(\hat{w}')=(1-q)d\hat{U}'(\hat{w}') . \tag{89}$$

Similar to $q$ at shrinkage, $q$ at swelling exists in the sample as well as the layer case (Eqs.(77a)-(77c) and (78a)-(78b)). It is worth noting that despite the same $q$ in Eqs.(76) and (89) at the corresponding soil water contents ($w'K=\hat{w}'\hat{K}$) the specific volume of the intra-aggregate matrix (per unit mass of the soil solids) at swelling ($\hat{U}'(\hat{w}')$) is less than that at shrinkage ($U'(w')$) because of the aggregate destruction in the beginning of the swelling stage and additional-interface-layer formation (see Sections 4.2.1 and 4.2.3).

One can express the specific crack volume at soil swelling, $\hat{U}_{cr}(\hat{w}')$ (see in Fig.5 the difference, $\hat{U}_{cr}-U_s=\hat{Y}-\hat{Y}_r$) through $\hat{U}(w)$ (for $\hat{U}(w)$ see the end of Section 3.2) as

$$\hat{U}_{cr}(\hat{w}')=U_s+q(U_h-U_z)/K-q(\hat{U}(w)-U_z)/\hat{K}, \qquad 0\leq \hat{w}'\leq \hat{w}'_h, \qquad 0\leq w\leq w_h \tag{90}$$

(cf. Eq.(80) at shrinkage). This presentation of $\hat{U}_{cr}(\hat{w}')$ is obtained after integrating the first Eq.(89), taking into account Eq.(52), the boundary condition from Eqs.(87) and (82), and $\hat{U}_z=U_z$ (see Fig.3) as well as the fact that $q$ does not depend on water content. Since $\hat{U}(w=0)\equiv\hat{U}_z=U_z$, Eq.(90) is reduced at $w=0$ to Eqs.(87) and (82), as it should be. Since $\hat{U}(w=w_h)\equiv\hat{U}_h=U_h$ (see Fig.3), Eq.(90) gives at $w=w_h$ the *specific* *residual volume of non-totally closed cracks after the shrink-swell cycle,* $\hat{U}_{cr}(\hat{w}'_h) \equiv \hat{U}_{crh}$ (Fig.5) to be

$$\hat{U}_{crh} = U_s + q(U_h - U_z)(1/K - 1/\hat{K}) . \tag{91}$$

Since $q>0$, $U_h>U_z$, and $K<\hat{K}$, according to Eq.(91), after a single (maximum) shrink-swell cycle $\hat{U}_{crh}>0$. Thus, Eq.(91) gives the *physical quantitative* explanation of the existence and origin of the non-total crack closing at swelling or, in other words, non-zero residual crack volume. The true physical reason for that is the existence of the interface layer and increase in its volume ($U_i \to \hat{U}_i$) at aggregate destruction (Sections 4.2.1 and 4.2.3). Indeed, $\hat{U}_i - U_i > 0$ leads to $\hat{K} > K$ (see Eq.(54)). Substituting for $\hat{U}'(\hat{w}')$ and $\hat{U}_{cr}(\hat{w}')$ in Eq.(86) their expressions from Eq.(52) and (90), we also obtain the useful presentation of the specific soil volume at swelling, $\hat{Y}(\hat{w}')$ (Fig.5) through $\hat{U}(w)$ as (cf. Eq.(81))

$$\hat{Y}(\hat{w}') = q(U_h - U_z)/K + (1-q)\hat{U}(w)/\hat{K} + qU_z/\hat{K} + \hat{U}_i + U_s , \quad 0 \leq \hat{w}' \leq \hat{w}'_h, \quad 0 \leq w \leq w_h . \tag{92}$$

Let us estimate the *decrease in maximum specific swelling volume after the shrink-swell cycle of an aggregated soil with cracks*, $\Delta Y_h \equiv Y_h - \hat{Y}_h$ (Fig.5). Denoting $\hat{Y}_h \equiv \hat{Y}(\hat{w}'_h)$, $\hat{Y}_{rh} \equiv \hat{Y}_r(\hat{w}'_h)$, it follows directly from Fig.5 that

$$\hat{Y}_h - \hat{Y}_{rh} = \hat{U}_{crh} - U_s . \tag{93}$$

From Eq.(93), $Y_h = U_{ah} + U_s$, and $\hat{Y}_{rh} = \hat{U}_{ah} + U_s$ (for $U_{ah}$ and $\hat{U}_{ah}$ see Fig.3), we have

$$\Delta Y_h \equiv Y_h - \hat{Y}_h = Y_h - \hat{Y}_{rh} - (\hat{U}_{crh} - U_s) = (U_{ah} - \hat{U}_{ah}) - (\hat{U}_{crh} - U_s) . \tag{94}$$

Thus, the maximum-specific-swelling-volume decrease, $\Delta Y_h$ includes the major contribution, $U_{ah} - \hat{U}_{ah}$ of the *reference* shrink-swell cycle (Section 4.2.3; Eq.(58b)) and the correcting (negative) contribution, $-(\hat{U}_{crh} - U_s)$ (Fig.5) of the residual volume of the non-totally closed cracks after the shrink-swell cycle. Substituting for $U_{ah} - \hat{U}_{ah}$ its expression from Eqs.(53) and (58b), and for $\hat{U}_{crh} - U_s$ its expression from Eq.(91), we finally obtain $\Delta Y_h$ to be

$$\Delta Y_h = (1-q)(U_h - U_z)(1/K - 1/\hat{K}) . \tag{95}$$

In the case of the reference shrink-swell cycle (i.e., for small samples) $q \to 0$ and $\Delta Y_h \to \Delta Y_{rh} \equiv Y_h - \hat{Y}_{rh} = U_{ah} - \hat{U}_{ah}$ (see Fig.5; cf. Eqs.(53) and (58b)). Thus, Eq.(95) gives the *physical quantitative* explanation of the observed effect of the decrease in maximum swelling volume, $\Delta Y_h > 0$ (e.g., [1]). The physical reason for this is the same as that above for the residual crack volume (Eq.(91)).

Replacing in Eqs.(90) and (92) $\hat{U}(w)/\hat{K}$ from Eqs.(51) and (52) one can present the specific crack volume at swelling, $\hat{U}_{cr}(\hat{W})$ and the swelling curve with cracks, $\hat{Y}(\hat{W})$ in different forms, through the *reference* swelling curve, $\hat{Y}_r(\hat{W})$ (Fig.5) and the $q$ factor as (cf. Eqs.(84) and (85) at shrinkage)

$$\hat{U}_{cr}(\hat{W}) - U_s = q(U_h - U_z)/K + qU_z/\hat{K} - q(\hat{Y}_r(\hat{W}) - \hat{U}_i - U_s) =$$
$$= q(U_h - U_z)/K - q(\hat{Y}_r(\hat{W}) - \hat{Y}_{rz}) = q(U_h - U_z)/K - q(\hat{Y}_r(\hat{W}) - Y_{rz}), \quad 0 \leq \hat{W} \leq \hat{W}_h \quad (96)$$

$$\hat{Y}(\hat{W}) = \hat{Y}_r(\hat{W}) + \hat{U}_{cr}(\hat{W}) - U_s = (1-q)\hat{Y}_r(\hat{W}) + q(U_h - U_z)/K + q\hat{Y}_{rz} =$$
$$= (1-q)\hat{Y}_r(\hat{W}) + q(U_h - U_z)/K + qY_{rz}, \quad 0 \leq \hat{W} \leq \hat{W}_h \quad (97)$$

($\hat{Y}_{rz} \equiv \hat{Y}_r(0) = U_z/\hat{K} + \hat{U}_i + U_s$; $\hat{Y}_{rz} = Y_{rz}$, see Fig.5). In Eqs.(96) and (97) $\hat{Y} \equiv \hat{Y}_s$, $\hat{U}_{cr} \equiv \hat{U}_{crs}$, and $q \equiv q_s$ in the sample case, or $\hat{Y} \equiv \hat{Y}_l$, $\hat{U}_{cr} \equiv \hat{U}_{crl}$, and $q \equiv q_l$ in the layer case. It can be checked that $\hat{Y}_z \equiv \hat{Y}(0)$ from Eq.(97) coincides with $Y_z = Y(0)$ as it should be (see Fig.5). Finally, the initial slope of the swelling curve with crack contribution from Eq.(97), $d\hat{Y}/d\hat{W}|_{\hat{W}=0} = (1-q)d\hat{Y}_r/d\hat{W}|_{\hat{W}=0}$ (Fig.5), similar to that of the reference swelling curve, $d\hat{Y}_r/d\hat{W}|_{\hat{W}=0}$ (see the end of section 4.2.4), depends on all the input parameters (see the end of section 4.1) and, in addition, on the initial sample size or layer thickness, and is only numerically found (Sections 6 and 7). It is worth reiterating that all the above results (Eqs.(90)-(97)) rely on the concept of the aggregate surface layer (or interface layer) and its change because of aggregate destruction at swelling.

In conclusion to the theoretical part the peculiarity of the swelling curve with cracks, $\hat{Y}(\hat{W})$ (Fig.5) should be noted. At any water content in the range, $0 \leq \hat{W} \leq \hat{W}_h$ the volume difference, $\hat{Y}(\hat{W}) - \hat{Y}_r(\hat{W})$ is stipulated by the crack volume, $\hat{U}_{cr}(\hat{W}) - U_s$ (Fig.5). Since any crack size is large compared to the size of the intra-aggregate clay pores, it is obvious that the soil cracks in the $0 \leq \hat{W} \leq \hat{W}_h$ range (by its definition as the range of contributive-clay swelling) stay empty and can only be filled in water in the range, $\hat{W}_h \leq \hat{W} \leq \hat{W}_h*$ (Fig.5; the *horizontal* part of the $\hat{Y}(\hat{W})$ curve) between the intersections of the saturation line with $\hat{Y}_r(\hat{W})$ and then with $\hat{Y}(\hat{W})$.

# 6. Analysis of available data

The objective of the data analysis is to give the *primary* experimental substantiation of the swelling model of an aggregated soil.

## *6.1. Data required*

Since we consider the physical, but not fitting, model, the data of two types are necessary for substantiation of the model: (i) input data for the model prediction of the (maximum) shrink-swell cycle; and (ii) the data for comparison with the model prediction to check the latter. According to [17], the *input* data include those on (i) initial soil sample height, $h$ (it is desirable that sample diameter $d \cong h$) or soil layer thickness, $h$ before shrinkage; (ii) soil solid density, $\rho_s$; *texture*: $c$ being clay content, $s_1$ being silt content (sand content, $s_2 = 1 - c - s_1$), and $x_m$ being the maximum sand grain size; and *initial structure*: $X_{mz}$ being the maximum aggregate size at the shrinkage limit and $P_z$ being the structural porosity at the shrinkage limit; and (iii) soil *shrinkage*: $Y_z$ being the specific volume at the shrinkage limit, in general, with cracks (Fig.5); $W_h$ being the water content at maximum swelling before shrinkage (Fig.5); and $U_{lph}$ being the specific volume of lacunar pores at maximum swelling before shrinkage (or $Y_h$ in Fig.5 instead of $U_{lph}$). In general, the minimum data for *comparison* with the model prediction should obviously include the data on the soil shrinkage and swelling curves with crack contribution. These data simultaneously include those on $\Delta Y_h$ (Fig.5). Also desirable are the data on the *reference* shrinkage and swelling curves for the soil, if those are available.

*6.2. Data used*

The data on swelling after shrinkage for a soil *layer* are lacking. We could only find the similar data that were obtained using soil *samples* [1]. These data relate to the samples with cracks (i.e., to sufficiently large samples), however, weakly differ from the data on reference shrinkage-swelling (see below). Peng and Horn [1] experimentally observed the shrink-swell behavior of two organic soils and two soils with a small content of organic matter. We only use data on the two latter soils since the model under consideration relates to the broad class of inorganic soils with the specific structure shown in Fig.1. These researchers [1] used the undisturbed soil cores with height, $h$=6.1cm and diameter, $d$=10 cm. They presented the data on shrinkage-swelling in the form of $e(\vartheta)$ (void ratio vs. moisture ratio for soil as a whole at shrinkage) and $\hat{e}(\hat{\vartheta})$ (void ratio vs. moisture ratio for soil as a whole at swelling). For this reason, predicting the shrinkage and swelling curves based on the above theory, we first transform the ($\vartheta$,$e$) variables to ($W$,$Y$) ones for shrinkage and ($\hat{\vartheta}$,$\hat{e}$) to ($\hat{W}$,$\hat{Y}$) for swelling and then, after predictions, return to ($\vartheta$,$e$) and ($\hat{\vartheta}$,$\hat{e}$) presentation of the predicted shrinkage and swelling curves ($W$=($\rho_w$/$\rho_s$)$\vartheta$, $Y$=($e$+1)/$\rho_s$, $\hat{W}$=($\rho_w$/$\rho_s$)$\hat{\vartheta}$, $\hat{Y}$=($\hat{e}$+1)/$\rho_s$). In particular, the moisture ratio, $\vartheta_h$ (Figs.6 and 7) is replaced with the corresponding $W_h$ (Fig.5) and void ratio, $e_z$ (Figs.6 and 7) with the corresponding $Y_z$ (Fig.5) as input values (see below).

Let us consider the *input* data. The soil $\rho_s$ values that are necessary for the prediction are not given in [1] (these values are not needed at the ($\vartheta$,$e$) presentation). For both soils we chose $\rho_s$=2.7 kgdm$^{-3}$ (from the physically possible range, 2.6-3 kgdm$^{-3}$). This choice does not influence the prediction in ($\vartheta$,$e$) and ($\hat{\vartheta}$,$\hat{e}$) presentation. The input data for the model prediction (except for the initial sample height) from [1] (plus $\rho_s$ as well as $W_h$ and $Y_z$ corresponding to $\vartheta_h$ and $e_z$) are presented in Table 1. We took the clay particles (illite) with clay content, $c$, silt and sand grains with content, $s_1$ and $s_2$ are in the usually accepted size ranges, 0-2μ, 2-50μ, and 50μ-$x_m$, respectively. The exact values for $x_m$ and $X_{mz}$ are lacking in [1]. Estimating the $x_m$ and $X_{mz}$ values that were used as input in predictions (Table 1) is discussed in section 6.3. We accepted the structural porosity, $P_z$ for both the soils under consideration to be $P_z$=0 (Table 1) since the experimental shrinkage curves (Figs.6 and 7, white squares) do not have the horizontal part at the intersection with the saturation line (cf. [17]). The lacunar pore volume at maximum swelling, $U_{lph}$ for both the soils under consideration was accepted as $U_{lph}$=0 (Table 1) since the experimental shrinkage curve points at $\vartheta$=$\vartheta_h$ (Figs.6 and 7, white squares; or $W$=$W_h$ in Fig.5) are on the saturation line (cf. [17]). Finally, the experimental values of the decrease in maximum swelling volume, $\Delta e_h^{exp}$ (in ($\vartheta$,$e$) variables) corresponding to $\Delta Y_h$ in Fig.5, after the shrink-swell cycle, were found from the experimental values of $e_h^{exp}$ and $\hat{e}_h^{exp}$ (see Figs.6 and 7 and Table 2) in order to be used in the swelling curve prediction (Section 6.4). The data from [1] that are used for *comparison* with the model prediction are reproduced by white circles (swelling curves) in Figs.6 and 7.

*6.3. Estimating some soil structure characteristics from the shrinkage curve data*

To physically predict the shrinkage and swelling curves of the soil samples with cracks one should know, in particular, the maximum sizes of sand grains, $x_m$ and aggregates, $X_{mz}$ (Section 6.1), but these data are lacking (Section 6.2). For this reason we use the shrinkage curve data below for the two soils from [1] to estimate $x_m$ and $X_{mz}$, and then predict the swelling curves. To estimate $x_m$ and $X_{mz}$ we use the approach that has been applied to similar task in [17] after corresponding modification (see below). In the indicated work $X_{mz}$ was estimated for four soils from [7] as a fitting parameter by fitting the theoretical reference shrinkage curve, $Y_r(W)$ (at given $\rho_s$, $c$, $s_1$, $P_z$, $W_h$, $Y_{rz}$, and $U_{lph}$ from Section 6.1) to experimental points obtained for small samples. In connection with such $X_{mz}$ estimating two results from [17] are important: (i) for a given set of the experimental points the fitted shrinkage curve is single-valued, but the found $X_{mz}$ value can vary in a small range $\Delta X_{mz}$ with $\Delta X_{mz}/X_{mz} \sim 0.01$-$0.02$; and (ii) at any $X_{mz}$ value from the found range the physically realized $x_m$ values can also be in a small range, $\Delta x_m \sim 5$-$15\mu$ around the mean $x_m$ value and lead to the same fitted reference shrinkage curve. Thus, $X_{mz}$ and $x_m$ of a particular soil can be estimated with accuracy, $\Delta X_{mz}$ and $\Delta x_m$. Correspondingly, the structural soil parameters, $k$ [24] and $K$ [25], depending on $X_{mz}$ and $x_m$, also have some small spreading [17] that does not influence the predicted (fitted) reference shrinkage curve.

In the case under consideration, estimating $x_m$ remains the same, but the above way of the estimating $X_{mz}$ should be a little modified. Indeed, unlike the case from [17] where the data on the reference shrinkage curve, $Y_r(W)$ were used to find the $X_{mz}$ values, in the case under consideration we deal with the experimental data on the shrinkage curves, $Y(W)$ (after transformation from $e(\vartheta)$) of the samples with cracks. Hence, to find $X_{mz}$ by fitting to the experimental data one should predict $Y(W)$ for samples with cracks. $Y(W)$ is expressed through the reference shrinkage curve, $Y_r(W)$ and the crack factor, $q_s$ (Eq.(85) in the sample case). Equation (85) at $W=W_z$ gives the relation between the known $Y_z$ (Table 1) and unknown $Y_{rz}$ as

$$Y_z = (1-q_s)Y_{rz} + q_s Y_h . \tag{98}$$

In this relation the crack factor, $q_s$ depends on $X_{mz}$ and $Y_{rz}$ at given $\rho_s$, $c$, $s_1$, $P_z$, $W_h$, and $U_{lph}$ from Table 1 [17,25] as well as on the initial sample size, $h$. Thus, Eq.(98) is nonlinear relative to $Y_{rz}$. Varying $Y_{rz}<Y_z$ at a fixed $X_{mz}$ (and given $Y_z$) and calculating the corresponding $q_s(Y_{rz},X_{mz})$ value [17], we numerically solve Eq.(98) and find $Y_{rz}$ at a given $X_{mz}$. Then for each $X_{mz}$ and corresponding $Y_{rz}$ we find the reference shrinkage curve, $Y_r(W)$ (at given $\rho_s$, $c$, $s_1$, $P_z$, $W_h$, and $U_{lph}$) [17]. Finally, using $X_{mz}$ as a *single* fitting parameter (starting from the possible $X_{mz}=2$mm), in series calculating $X_m$ (Eq.(57) without the "^" sign [25]), $h^*$ (Eq.(79)), $q_s$ (Eq.(77)), the theoretical $Y(W)$ curve (Eq.(85)), and then $e(\vartheta)$ for a number of $X_{mz}$ values, and utilizing the usual least-square criterion, we fit the theoretical $e(\vartheta)$ to experimental $e^{exp}(\vartheta^{exp})$ (white squares in Fig.6 for soil 03 and Fig.7 for soil 04). The calculated $X_{mz}$ and $x_m$ are shown in Table 1 (relative to $x_m$ see the above remarks). The corresponding parameters of the soil texture, structure, cracking, and intra-aggregate structure during shrinkage, as well as fitting characteristics ($r_e^2$ and $\sigma_e$) are presented in Table 3. Table 4 indicates the intermediate soil characteristics (that will be needed below) found in the course of the calculations. The crack factor, $q_s$ for the two soils is so small (Table

3) that the predicted reference shrinkage curve, $e_r(\vartheta)$ practically coincides with the shrinkage curve for samples with cracks, $e(\vartheta)$ (curve 1) in the scale of Figs.6 and 7. Such approximate coincidence is expected since the slope of the experimental shrinkage curve (white squares in Figs.6 and 7) in the basic shrinkage range is approximately equal to unity. To additionally emphasize the approach possibilities, Figs.6 and 7 also show, for the two soils, the shrinkage curve of the layer of the same thickness, $h$=6.1cm, $e_l(\vartheta)$ (curve 3). The crack factor for the layer, $q_l$ for the found $h/h^*$ ratio is shown in Table 3.

### *6.4. Model prediction of swelling curves*

The model prediction of swelling curves with cracks for sample or layer relies on the estimates that are presented in Tables 3 and 4, and on the theory from sections 2-5. According to the theory, in addition to the soil parameters from Tables 3 and 4 one should know the $\hat{K}$ ratio at swelling: (i) immediately; (ii) through the interface layer volume at swelling, $\hat{U}_i$ (Eq.(54)); or (iii) through the maximum aggregate size at swelling after aggregate destruction, $\hat{X}_{mz}$ (Eq.(56)). There is no data on $\hat{X}_{mz}$ for the two soils. As the primary data we take advantage of the maximum-swelling-volume decrease, $U_{ah}-\hat{U}_{ah}$ (in terms of $(e,\vartheta)$ see $\Delta e_h^{exp}$ in Table 2) that enables one to estimate $\hat{U}_i$ (Eq.(58b)) and then $\hat{K}$ (Eq.(54)). The $\hat{U}_i$ and $\hat{K}$ estimates are shown in Table 5. In addition, using the value found for $\hat{U}_i$ as well as Eqs.(56) and (57), we estimated the maximum aggregate size after destruction at $\hat{W}$=0, $\hat{X}_{mz}$ and at maximum swelling, $\hat{X}_m$ (Table 5).

As in the case of shrinkage, the prediction of the swelling curve consists of two parts: the prediction of the reference swelling curve and, then, the prediction of the crack volume through the crack factor found, $q_s$ or $q_l$. As a whole, the general algorithm of the prediction procedure includes a number of steps. (i) For $P_h=P_z=0$ (Table 1), $v_s$, $v_z$, $v_h$, $u_s$, $u_S$, $u_z$, $U_h$, $U_z$, $U_i$ (Table 4), $k$, $K$, $q_s$, and $q_l$ (Table 3), and $\hat{K}$ (Table 5) one finds in series $\hat{U}(w)$ (Eqs.(32) and (27) after $u \to \hat{u}$ and $U \to \hat{U}$), $\hat{U}'(\hat{w}')$ (Eq.(52)), $\hat{U}_a(\hat{w}')$ (Eq.(50a)), $\hat{Y}_r(\hat{w}')$ (Eq.(51)), and then $\hat{Y}(\hat{w}')$ (Eq.(97)) for the sample (with $q_s$) and for the layer (with $q_l$). (ii) One solves Eq.(69) for $\hat{R}(\zeta)$ (Fig.4, curve 6). (iii) One finds $\Delta\hat{\omega}(\hat{w}')$ (Eq.(66)), $\omega(\hat{w}')$ (Eq.(63)), and $\hat{W}(\hat{w}')=\hat{w}'+\Delta\hat{\omega}(\hat{w}')+\omega(\hat{w}')$. (iv) $\hat{Y}_r(\hat{w}')$, $\hat{Y}(\hat{w}')$, and $\hat{W}(\hat{w}')$ determine the reference swelling curve, $\hat{Y}_r(\hat{W})$ and that with cracks, $\hat{Y}(\hat{W})$ for the sample and layer. (v) Finally, $\hat{Y}_r(\hat{W})$, $\hat{Y}(\hat{W})$ (for sample), and $\hat{Y}_l(\hat{W})$ (for layer) are recalculated to the predicted curves, $\hat{e}_r(\hat{\vartheta})$, $\hat{e}(\hat{\vartheta})$, $\hat{e}_l(\hat{\vartheta})$ in Figs.6 and 7.

In addition to the above *physical* prediction of the swelling curves, $\hat{e}_r(\hat{\vartheta})$, $\hat{e}(\hat{\vartheta})$, $\hat{e}_l(\hat{\vartheta})$ (Figs.6 and 7) we used the same algorithm, regarding the physical parameter, $\hat{K}$ as a fitted one (but not as found from the $\Delta e_h^{exp}$ data of Table 2). Results of fitting the theoretical curve, $\hat{e}(\hat{\vartheta})$ for samples to the experimental swelling curve (white circles in Figs 6 and 7), are presented in Figs.6 and 7 (dashed curve) and in Table 5 (the $\hat{K}_f$ value). The corresponding minimum sum of squares, $\hat{\Sigma}_{min}$ of the differences between the experimental values, $\hat{e}_i^{exp}$ (i=1,…, $M$; $M$ being the number of white circles in Fig.6 or Fig.7) and the corresponding predicted $\hat{e}(\hat{\vartheta}_i^{exp})$ values, allows one to estimate the standard deviations, $\sigma_{\hat{e}}$ of the $\hat{e}_i^{exp}$ values [28] as $\sigma_{\hat{e}} = (\hat{\Sigma}_{min}/(M-1))^{1/2}$ (Table 5) and the goodness of fit, $r_{\hat{e}}^2$ (Table 5) of the theoretical $\hat{e}(\hat{\vartheta})$ curve in Figs.6 and 7. Using $\sigma_{\hat{e}}$ (Table 5), besides the mean $\hat{K}_f$ values (Table 5), we estimated their standard errors, $\Delta \hat{K}_f$ (Table 5). To this end we estimated the best-fit $\hat{K}_f$ values for the experimental $\hat{e}_i^{exp}$ points (white circles in Figs.6 and 7) after their displacement up and down by the above found standard deviation, $\sigma_{\hat{e}}$. The $\Delta \hat{K}_f$ values will be used below. The results of the comparison between the above predictions and available experimental data are presented in Section 7.3.

# 7. Results and discussion
## 7.1. Theoretical results

The four major theoretical results are the development of the physical models of swelling curves for: (i) clay (section 2); (ii) intra-aggregate matrix (section 3); (iii) aggregated soil without cracks (section 4); and (iv) soil with cracks (section 5). The specific results are the expressions for: (1) the shape of the clay swelling curve; (2) the slope of the curves for the clay and intra-aggregate matrix at the start of swelling; (3) the maximum-swelling-volume decrease of an aggregated soil after the reference shrink-swell cycle (Eq.(58b)); (4) the swelling curve of an aggregated soil with cracks through the reference swelling curve and characteristics of previous shrinkage (Eq.(97)); (5) the crack volume of an aggregated soil at swelling through the reference swelling curve and characteristics of previous shrinkage (Eq.(96)); (6) the residual crack volume of the aggregated soil after the maximum shrink-swell cycle (Eq.(91)); and (7) the maximum-swelling-volume decrease of the aggregated soil after the maximum shrink-swell cycle, accounting for both the contribution of the reference shrinkage-swelling and that of the residual cracks (Eq.(95)). All these results have a quantitative form and are in the agreement with observations [1].

The indicated results show: (a) the fundamental possibilities of the physical approach and (b) the qualitative experimental substantiation of the approach since the phenomena of residual cracking and decrease of the maximum swelling volume after the maximum shrink-swell cycle are experimentally observed [1]. Similar results can not be obtained by any fitting approach that only describes a set of experimental points by a multiple fitting, but not explains.

The major features of the modeling are:
1. The physical, rather than fitting prediction of swelling curves from a number of soil parameters that can be measured independently of the experimental swelling curve.
2. The close interconnection between the above four models. We show step-by-step transition from contributive-clay swelling to soil swelling through the intra-aggregate matrix swelling and reference swelling based on soil texture, structure, and new concepts of the intra-aggregate structure, interface layer and lacunar pores [13-15], and the structure variation at the beginning of swelling as a result of air entrapping and aggregate destruction. Each the following model generalizes the previous.
3. The four above developed swelling curve models are closely connected, and not only with each other. Each swelling model (of clay, intra-aggregate matrix, soil without and with cracks) is the generalization of the corresponding available shrinkage model. In the course of such a generalization for each pair of shrink-swell "twins", the similarities and differences between shrinkage and swelling are emphasized. In each case the shrinkage concepts are indicated that: (i) are not suitable in the swelling case; (ii) can be accepted after some modification; and (iii) are retained at swelling. Such an approach permits us to use and rely on the construction methodology of the above physical models of shrinkage.

### *7.2. Model checking as applied to shrinkage curves*

In section 6.3 we used fitting of the theoretical $e(\vartheta)$ curve to the experimental data, $e^{exp}(\vartheta^{exp})$ (white squares in Figs.6 and 7) with one fitted parameter, $X_{mz}$ since there are no $X_{mz}$ and $Y_{rz}$ data for the soils under consideration. The corresponding minimum sum of squares, $\Sigma_{min}$ of the differences between the experimental values, $e_i^{exp}$ (i=1,..., $N$; $N$ being the number of white squares in Fig.6 or Fig.7) and the corresponding predicted $e(\vartheta_i^{exp})$ values, allows one to estimate the standard deviations, $\sigma_e$ of the $e_i^{exp}$ values [28] as $\sigma_e=(\Sigma_{min}/(N-1))^{1/2}$ (Table 3) and the goodness of fit, $r_e^2$ (Table 3) of the theoretical $e(\vartheta)$ curve in Figs.6 and 7. The comparison in Figs.6 and 7 between $e(\vartheta)$ and experimental data (white squares) using the $r_e^2$ and $\sigma_e$ values, shows that the predicted shrinkage curves, $e(\vartheta)$ (for soils 03 and 04) are in agreement with the experimental data not only from the viewpoint of the fitting criterion, connected with the high $r_e^2$, but also from the viewpoint of the standard physical criterion - the difference between the predicted $e(\vartheta)$ value at $\vartheta=\vartheta_i^{exp}$ and corresponding experimental value, $e_i^{exp}$ does not exceed the two standard deviations, $|e(\vartheta_i^{exp})-e_i^{exp}|<2\sigma_e$ for each i=1,..., $N$ (see $\sigma_e$ in Table 3). In the context of the model substantiation as applied to the shrinkage curve prediction, it is worth noting the $X_{mz}$, $k$, $q_s$, and $K$ values (Table 3). The $X_{mz}$ values found are physically reasonable (usually up to the first millimeters; cf. [17]). The $k$ and $q_s$ found (Table 3) give the values of the shrinkage curve slope, $S$ in the basic shrinkage range (Eq.(83); the slopes in $(W,Y)$ and $(\vartheta,e)$ coordinates numerically coincide) for soils 03 and 04 (see Figs.6 and 7; curves 1 and 3 and the captions), that are physically reasonable and correspond to the visual estimate, $S\cong1$ for white squares. Finally, the found $K$ values (Table 3) are in agreement with another estimate, $K=W_h/w'_h$ based on reference shrinkage [13,14] (see $W_h$ and $w'_h$ in Fig.3). Substantiation of the model as applied to the shrinkage curve has been conducted in earlier works [13,14,17]. The above fitting results additionally show evidence in favor of the shrinkage model.

### *7.3. Model checking as applied to swelling curves*

The fitted swelling curve of the samples (the dashed curve in Figs.6 and 7) is in good agreement with the data (white circles) from the viewpoint of both the fitting criterion (high $r_{\hat{e}}^2$ in Table 5) and the standard physical criterion, since the deflections of the fitted curve of the experimental points (white circles) are within the limits of the two standard deviations, $2\sigma_{\hat{e}}$ (Table 5). To check the physically predicted swelling curve of the samples (solid curve 2 in Figs.6 and 7) we can compare this curve with the fitted one (the dashed curve). In addition, we can compare the aggregate/intra-aggregate mass ratios after aggregate destruction, as found for these curves, $\hat{K}$ and $\hat{K}_f$, accounting for the standard deviations, $\Delta\hat{K}_f$ (Table 5). One can see from Table 5 that $|\hat{K}-\hat{K}_f|<\Delta\hat{K}_f$. Moreover, the difference between the physically predicted and fitted swelling curves (between curve 2 and the dashed one in Figs.6 and 7) does not exceed the two standard deviations, $2\sigma_{\hat{e}}$ (Table 5). Thus, the physically predicted swelling curve for the sample case, $\hat{e}(\hat{\vartheta})$ for soils 03 and 04 in Figs.6 and 7, respectively, is in agreement with the available experimental data. This result gives primary validation of the model prediction. In addition, this result justifies the preliminary assumption in frame of the model relative to the swelling behavior of soils 03 and 04 from [1] (see Section 2.2; Eq.(15) and the paragraph after Eq.(15)).

Thus, the quantitative physical prediction of a swelling curve is possible according to Sections 7.2 and 7.3 and Figs.6 and 7 if there are usual input data as in Table 1. This prediction do not requires preliminary measurements of swelling curve points for the following fitting approximation. Similar results relative to the practical applications of the physical prediction of a shrinkage curve are available in [13-15,17] where the data on more that 20 soils were used.

The estimated maximum aggregate sizes after destruction, $\hat{X}_{mz}$ and $\hat{X}_m$ (Table 5) are essentially smaller than those before destruction, $X_{mz}$ and $X_m$ (Table 3). Such strong aggregate destruction is likely to be connected with the consideration of the *maximum* shrink-swell cycle when the water content varies in the *maximum* ranges, $0<W<W_h$ and $0<\hat{W}<\hat{W}_h$. In real conditions the water content usually varies in the relatively small ranges, $\Delta W<<W_h$ and $\Delta \hat{W}<<\hat{W}_h$, and the aggregate destruction ($X_{mz} \rightarrow \hat{X}_{mz}$) should occur appreciably more slowly.

## *7.4. Illustrative results*

A number of results obtained after calculations and intermediate ones obtained in the course of calculations cannot be compared with relevant data since too much data are missing. These results are presented by the values of different soil and sample parameters in Tables 2-5 and by the different curves that are characteristic for the soils, in Figs.6-10. These results are interesting, at least as illustrative ones, indicating to the reader that is interested, the order of magnitude of parameter values and mutual arrangement of the curves. The objective of this section is a short comment on the content (or a part of it) of Figs.6-10 and Tables 2-5. Figures 6 and 7, besides the shrinkage ($e(\vartheta)$) and swelling ($\hat{e}(\hat{\vartheta})$) curves for the sample ($h$=6.1cm) with cracks (curves 1 and 2) for which there are data (white squares and circles), also show the shrink-swell cycle for the layer of the same thickness ($h$=6.1cm), $e_l(\vartheta)$ and $\hat{e}_l(\hat{\vartheta})$ (curves 3 and 4; note that the fitting models of shrinkage [4-12] and swelling [1] curves, unlike the physical model under consideration, do not regard the layer case and only relate to the sample case). Figures 6 and 7 emphasize the essential difference between these two shrink-swell cycles (for samples and layers) and simultaneously show interconnections between them through the reference shrink-swell cycle and crack volume (as a function of water content) that is different for the sample case and the layer one. For these soils in the sample case the crack volume is very small (see the $q_s$ value in Table 3). Figures 8 and 9 illustrate for the particular real soils the shrink-swell cycle of the contributive clay (paste) and intra-aggregate matrix, respectively. Figure 10 illustrates for the particular soil the maximum-internal-size dependence of the water-filled pores in the contributive clay on the water content at swelling, $\hat{R}(\zeta)$. $\hat{R}(\zeta)$ determines the contribution, $\hat{\omega}(\zeta)$ of the interface layer to the soil water content at swelling (cf. the similar dependence $R(\zeta)$ at shrinkage in Fig.4, curve 3). Figures 6 and 7 show that the residual crack volume in the layer case is more than in the sample case: $(\hat{e}_{crh}-e_s)_l > (\hat{e}_{crh}-e_s) \sim 0$ (for $(\hat{e}_{crh}-e_s)_l$ see also Table 2). On the contrary, the decrease in maximum swelling volume after the soil shrink-swell cycle in the layer case is less than in the sample case: $(\Delta e_h)_l < \Delta e_h^{exp}$ (see Table 2 as well as Figs.6 and 7). These results are physically expected. Finally, the comparison between the parameters from Table 5 and similar ones, but without the "∧" sign, from Tables 3 and 4, illustrates the variation of these parameters as a result of the aggregate destruction in the beginning of the swelling stage. For each parameter this variation corresponds to the behavior indicated in section 4.2.1. For instance, $\hat{X}_{mz} < X_{mz}$, $\hat{K} > K$, $\hat{U}_i > U_i$.

## 8. Conclusion

The aim of this work is the physical modeling of the soil swelling curve after previous shrinkage in the total range of water content that is possible for free shrinkage and swelling (without loading). Eventually, we consider one maximum shrink-swell cycle of an aggregated soil. The methodology is based on the construction of the physical-swelling-curve chain for clay, intra-aggregate matrix, and aggregated soil without and with cracks, relying on the available shrinkage models for these soil media [13-17,19,20,23,25] with necessary modifications. All the major concepts that were used in the available shrinkage models: aggregate surface layer (or interface layer) and its specific volume, aggregate/intra-aggregate mass ratio, lacunar factor, critical sample/layer size, and crack factor, are subject to relevant modification and comprehension as applied to the swelling case. The model predictions are based on the soil texture and structure (including the intra-aggregate), sample size/layer thickness, and several physical soil characteristics that also determine the reference shrinkage curve and are measured by known means [17]. The model physically predicts such observable peculiarities as the residual crack volume and decrease in maximum soil volume after the shrink-swell cycle. The primary experimental confirmation of the model and model concepts is reached by the analysis of the relevant data [1] on the shrink-swell cycle of two inorganic soils with crack development. The limitedness of the reliable experimental data that are relevant to the aims of this work, i.e., the data on the shrink-swell cycles of aggregated soils, is obvious. Therefore, additional experimental checking is desirable. Nevertheless, the results of the above analysis are promising.

Results of this work create prerequisites of the theoretical insight into: (i) multifold soil shrinkage-swelling with cracking, arbitrary shrinkage and swelling ranges within the limits of the maximum ($0 \leq W \leq W_h$ and $0 \leq \hat{W} \leq \hat{W}_h$), and possible soil loading (overburden pressure); as well as (ii) the impact of shrink-swell cycles on the actually observed crack network, crusting, hydraulic properties, water flow, and transport phenomena.

**Notation**

$c, c_*$    soil clay content and its critical value (dimensionless)
$e, \hat{e}$    void ratio of clay matrix or soil at shrinkage and swelling (dimensionless)
$F, \hat{F}$    saturation degree of clay matrix at shrinkage and swelling (dimensionless)
$F(\eta,\beta)$ aggregate-size distribution at structural porosity, $\beta$ (dimensionless)
$F_i$    volume fraction of water-filled interface clay pores (dimensionless)
$F_z$    $F$ value at $\zeta=\zeta_z$ (dimensionless)
$f(\eta,\hat{P})$ clay pore-size distribution at porosity, $\hat{P}$ for swelling (dimensionless)
$G(\alpha,\beta,\chi)$      function from Eq.(40) (dimensionless)
$K, \hat{K}$    aggregate/intra-aggregate mass ratio at *shrinkage* and *swelling* (dimensionless)
$\hat{K}_f$    $\hat{K}$ ratio estimated as fitting parameter (dimensionless)
$k$    lacunar factor (dimensionless)
$P(\zeta), \hat{P}(\zeta)$     clay matrix porosity at shrinkage and swelling (dimensionless)
$P_h, P_z$ structural porosity at shrinkage close to $W=W_h$ and $W=0$ (dimensionless)

$\hat{P}_z$ structural porosity at swelling close to $\hat{W}=0$ (dimensionless)

$p$ silt and sand porosity in the state of *imagined* contact (dimensionless)

$R(w')$ maximum internal size of water-filled clay pores at shrinkage (μm)

$R_m$ maximum size of clay pores in interface layer part of the $U_i$ volume (μm)

$R_{min}$ minimum size of clay pores in interface layer part of the $U_i$ volume (μm)

$\hat{R}(\hat{w}')$ maximum internal size of water-filled clay pores at swelling (μm)

$\hat{R}_m$ maximum size of clay pores in interface layer part of the $\Delta\hat{U}_i$ volume (μm)

$\hat{R}_{min}$ minimum size of clay pores in interface layer part of the $\Delta\hat{U}_i$ volume (μm)

$r_m(\zeta), r_o(\zeta)$ maximum and minimum internal size of clay pores at shrinkage (μm)

$r_{mM}$ maximum external size of clay pores at $\zeta=1$ (μm)

$\hat{r}_m(\zeta), \hat{r}_o(\zeta)$ maximum and minimum internal size of clay pores at swelling (μm)

$r_e^2, r_{\hat{e}}^2$ goodness of fit of sample shrinkage and swelling curve (dimensionless)

$U, \hat{U}$ specific volume of intra-aggregate matrix at shrinkage and swelling (dm$^3$kg$^{-1}$)

$U_a, \hat{U}_a$ specific volume of aggregates at shrinkage and swelling (dm$^3$kg$^{-1}$)

$U_{ah}$ $U_a$ value at maximum swelling of intra-aggregate matrix (dm$^3$kg$^{-1}$)

$U_{cr}, \hat{U}_{cr}$ specific crack volume at shrinkage and following swelling (dm$^3$kg$^{-1}$)

$U_{crz}$ $U_{cr}$ value at the end of shrinkage (dm$^3$kg$^{-1}$)

$U_h$ $U$ value at $W=W_h$ (dm$^3$kg$^{-1}$)

$U_i, \hat{U}_i$ interface layer contribution to the specific volume of aggregates at shrinkage and swelling, respectively (dm$^3$kg$^{-1}$)

$U_s$ specific volume of structural pores (dm$^3$kg$^{-1}$)

$U_z$ $U$ value in the oven-dried state (dm$^3$kg$^{-1}$)

$U', \hat{U}'$ intra-aggregate matrix contribution to the specific volume of aggregates at shrinkage ($U'=U/K$) and swelling ($\hat{U}'=\hat{U}/\hat{K}$), respectively (dm$^3$kg$^{-1}$)

$\hat{U}_{ah}$ $\hat{U}_a$ value at maximum swelling of intra-aggregate matrix (dm$^3$kg$^{-1}$)

$\hat{U}_{az}$ $\hat{U}_a$ value in the oven-dried state of intra-aggregate matrix (dm$^3$kg$^{-1}$)

$\hat{U}_{crh}$ $\hat{U}_{cr}$ value at the end of shrink-swell cycle with cracking (dm$^3$kg$^{-1}$)

$\hat{U}_{cr\,l}, \hat{U}_{cr\,s}$ $\hat{U}_{cr}$ value in case of soil layer and sample, respectively (dm$^3$kg$^{-1}$)

$\hat{U}_{crz}$ $\hat{U}_{cr}$ value at the start of soil swelling after shrinkage with cracking (dm$^3$kg$^{-1}$)

$\hat{U}_z$ $\hat{U}$ value in the oven-dried state (dm$^3$kg$^{-1}$)

$u(\zeta)$ relative volume of soil intra-aggregate matrix at shrinkage (dimensionless)

$u_h, \hat{u}_h$ $u$ and $\hat{u}$ values at the maximum swelling point ($u_h=\hat{u}_h$) (dimensionless)

$u_S, \hat{u}_S$ relative volume of non-clay solids of intra-aggregate matrix at shrinkage and swelling ($u_S=\hat{u}_S$) (dimensionless)

$u_s, \hat{u}_s$ relative volume of solid phase of intra-aggregate matrix at shrinkage and swelling ($u_s=\hat{u}_s$) (dimensionless)

$u_z, \hat{u}_z$ $u$ and $\hat{u}$ values at $\zeta=0$ ($u_z=\hat{u}_z$) (dimensionless)

$u_{cp}(\zeta)$ relative volume of clay matrix pores in soil intra-aggregate matrix at shrinkage (dimensionless)

$u_{lp}(\zeta), \hat{u}_{lp}(\zeta)$ relative volume of lacunar pores in soil intra-aggregate matrix at shrinkage and swelling, respectively (dimensionless)

$u_{lph}, u_{lpz}$ $u_{lp}$ values at $\zeta=\zeta_h$ and $\zeta=\zeta_z$, respectively (dimensionless)

$\hat{u}(\zeta)$ relative volume of soil intra-aggregate matrix at swelling (dimensionless)

$\hat{u}_{lph}, \hat{u}_{lpz}$ $\hat{u}_{lp}$ values at $\zeta=\zeta_h$ ($\hat{u}_{lph}=u_{lph}$) and $\zeta=\zeta_z$ ($\hat{u}_{lpz}=u_{lpz}$) (dimensionless)

$V, \hat{V}$ specific volume of clay matrix at shrinkage and swelling (dm$^3$kg$^{-1}$)

$v, \hat{v}$ ratio of clay volume at shrinkage and swelling to possible clay volume maximum in the solid state (the liquid limit) (dimensionless)

$v_h, v_n, v_z$ $v$ value at $\zeta=\zeta_h$, $\zeta=\zeta_n$, and $\zeta=\zeta_z$, respectively (dimensionless)

$v_s, \hat{v}_s$ relative clay solid volume at shrinkage and swelling ($v_s=\hat{v}_s$) (dimensionless)

$\hat{v}_h, \hat{v}_z$ $\hat{v}$ value at $\zeta=\zeta_h$ ($\hat{v}_h=v_h$) and $\zeta=\zeta_z$ ($\hat{v}_z=v_z$) (dimensionless)

$W, \hat{W}$ total water content of soil at shrinkage and swelling (kg kg$^{-1}$)

$W_h$ $W$ value at shrinkage start (kg kg$^{-1}$)

$\hat{W}_h$ $\hat{W}$ value at swelling finish ($\hat{W}_h < W_h$) (kg kg$^{-1}$)

$\hat{W}_h*$ water content (Fig.5) at which residual cracks after shrink-swell cycle would be water-filled (kg kg$^{-1}$)

$w$ water content of soil intra-aggregate matrix at shrinkage (kg kg$^{-1}$)

$w_h, w_n$ $w$ value at $\zeta=\zeta_h$ ($w_h=\hat{w}_h$) and $\zeta=\zeta_n$ ($w_n=\hat{w}_n$), respectively (kg kg$^{-1}$)

$\hat{w}$ water content of soil intra-aggregate matrix at swelling ($\hat{w}=w$) (kgkg$^{-1}$)

$\overline{w}, \hat{\overline{w}}$ water content of clay matrix at shrinkage and swelling ($\overline{w}=\hat{\overline{w}}$) (kg kg$^{-1}$)

$\overline{w}_h, \overline{w}_n$ $\overline{w}$ value at $\zeta=\zeta_h$ ($\overline{w}_h=\hat{\overline{w}}_h$) and $\zeta=\zeta_n$ ($\overline{w}_n=\hat{\overline{w}}_n$), respectively (kg kg$^{-1}$)

$w', \hat{w}'$ contribution of intra-aggregate matrix to total water content at shrinkage and swelling, respectively (kgkg$^{-1}$)

$w'_h$ $w'$ value at maximum swelling (shrinkage start) (kg kg$^{-1}$)

$w'_s$ $w'$ value at the end point of structural shrinkage (kg kg$^{-1}$)

$\hat{w}'_b$ $\hat{w}'$ value being the beginning point of water filling in the interface layer part of the $U_i$ volume at swelling (kg kg$^{-1}$)

$\hat{w}'_h$ $\hat{w}'$ value at swelling finish (kg kg$^{-1}$)

$\hat{w}'_e$ $\hat{w}'$ value being the end point of water filling in the interface layer part of the $\Delta\hat{U}_i$ volume at swelling (kg kg$^{-1}$)

$X_m, X_{mz}$ maximum aggregate size at shrinkage start and finish (mm)

$\hat{X}_{mz}, \hat{X}_m$ maximum aggregate size at swelling start and finish (mm)

$x_n$ mean size of clay particles and silt and sand grains of soil (μm)

$Y$ specific soil volume at shrinkage with cracking (dm$^3$kg$^{-1}$)

$Y_h$ $Y$ value at the start of shrinkage (maximum swelling) (dm$^3$kg$^{-1}$)

$Y_r$ specific soil volume at reference shrinkage (dm$^3$kg$^{-1}$)

$Y_{rh}$ specific soil volume at the start of reference shrinkage (at $W=W_h$) (dm$^3$kg$^{-1}$)

$Y_{rz}$ specific soil volume at the end of reference shrinkage (dm$^3$kg$^{-1}$)

$\hat{Y}$ specific soil volume at swelling after shrinkage with cracking (dm$^3$kg$^{-1}$)

$\hat{Y}_h$ $\hat{Y}$ value after shrink-swell cycle with cracking (dm$^3$kg$^{-1}$)

$\hat{Y}_l, \hat{Y}_s$ $\hat{Y}$ value in case of soil layer and sample (dm$^3$kg$^{-1}$)

$\hat{Y}_r$ specific soil volume at reference swelling (dm$^3$kg$^{-1}$)

$\hat{Y}_{rh}$ specific soil volume after reference shrink-swell cycle (dm$^3$kg$^{-1}$)

$\hat{Y}_{rz}$ specific soil volume at the start of reference swelling (dm$^3$kg$^{-1}$)

$\hat{Y}_z$ $\hat{Y}$ value at the start of soil swelling after shrinkage with cracking (dm$^3$kg$^{-1}$)

$\alpha, \beta, \chi$ parameters of $G$ function of Eq.(40) (dimensionless)

$\Delta\hat{K}_f$ standard deviation of the $\hat{K}_f$ value (dimensionless)

| | |
|---|---|
| $\Delta \hat{U}_i$ | additional interface layer volume that appears at swelling (dm$^3$kg$^{-1}$) |
| $\Delta Y_h$, $\Delta Y_{rh}$ | maximum-specific-swelling-volume decrease after shrink-swell cycle with cracking and reference shrink-swell cycle, respectively (dm$^3$kg$^{-1}$) |
| $\Delta \hat{\omega}$ | water contribution of the additional interface layer at swelling (kg kg$^{-1}$) |
| $\zeta$ | relative water content in clay or soil intra-aggregate matrix (dimensionless) |
| $\zeta_h$ | maximum swelling point on the $\zeta$ axis (dimensionless) |
| $\zeta_n$ | end point of basic shrinkage (the air-entry point) (dimensionless) |
| $\zeta_z$ | shrinkage limit on the $\zeta$ axis (dimensionless) |
| $\zeta_a$ | $\zeta$ value at shrinkage when an adsorbed film only remains (dimensionless) |
| $\zeta_e$, $\zeta_b$, $\zeta_s$ | correspond to $\hat{w}'_e$, $\hat{w}'_b$, and $w'_s$ on the $\zeta$ axis (dimensionless) |
| $\eta$ | parameter in Eqs.(41), (47), (55), (63), (64), (66), (70), (72) (dimensionless) |
| $\theta$ | moisture ratio of clay matrix (dimensionless) |
| $\Pi_h$ | clay porosity of interface layer part of the $U_i$ volume (dimensionless) |
| $\Pi_z$ | clay porosity of interface layer part of the $\Delta \hat{U}_i$ volume (dimensionless) |
| $\rho_s$ | density of clay solids or mean density of soil solids (kg dm$^3$) |
| $\rho_w$ | water density (kg dm$^3$) |
| $\sigma_e$ | standard deviations of experimental shrinkage curve values (dimensionless) |
| $\sigma_{\hat{e}}$ | standard deviations of experimental swelling curve values (dimensionless) |
| $\omega$ | interface layer contribution to the total water content at shrinkage (kgkg$^{-1}$) |
| $\omega_h$ | $\omega$ value at shrinkage start (kg kg$^{-1}$) |
| $\hat{\omega}$ | interface layer contribution to the total water content at swelling (kg kg$^{-1}$) |

**Figure captions**

**Fig.1.** Schematic illustration of the accepted soil structure [13,14,17]. Shown are (1) an assembly of many soil aggregates and inter-aggregate pores contributing to the specific soil volume, $Y$; (2) an aggregate as a whole contributing to the specific volume $U_a=U_i+U'$; (3) an aggregate with two parts indicated: (3a) an interface layer contributing to the specific volume $U_i$ and (3b) an intra-aggregate matrix contributing to the specific volumes $U$ and $U'=U/K$; (4) an aggregate with indicated intra-aggregate structure: (4a) clay, (4b) silt and sand grains, and (4c) lacunar pores; and (5) an inter-aggregate pore leading at shrinkage to an inter-aggregate crack contributing to the specific volume $U_{cr}$. $U$ is the specific volume of an intra-aggregate matrix (per unit mass of the oven-dried matrix itself). $U'$ is the specific volume of an intra-aggregate matrix (per unit mass of the oven-dried soil). $U_i$ is the specific volume of the interface layer (per unit mass of the oven-dried soil). $U_{cr}$ is the specific volume of cracks (per unit mass of the oven-dried soil). $U_a$ is the specific volume of aggregates (per unit mass of the oven-dried soil). $K$ is the aggregate/intra-aggregate mass ratio.

**Fig.2.** General qualitative view of clay shrinkage ($v(\zeta)$) and swelling ($\hat{v}(\zeta)$) curves in relative coordinates ($\zeta, v$) and ($\zeta, \hat{v}$). ($\zeta_z, v_z$), ($\zeta_n, v_n$), and ($\zeta_h, v_h$) are characteristic points of the shrinkage curve. $(0, v_z)$ and ($\zeta_h, v_h$) are characteristic points of the swelling curve. The inclined dash line going with slope, $1-v_s$ through point $(0, v_s)$ is the saturation line. The inclined dash line going with slope, $2(1+v_s-2v_z)$ through point $(0, v_z)$ is the initial tangent to the clay swelling curve. Depending on clay type (i.e., $v_s$ and $v_z$) $2(1+v_s-2v_z)$ can be more or less than $1-v_s$.

**Fig.3.** Illustrative scheme of transforming the shrink-swell cycle of an intra-aggregate matrix, $U(w)$ and $\hat{U}(\hat{w})$ (dash lines 1 and $\hat{1}$) to auxiliary curves, $U'(w')$ and $\hat{U}'(\hat{w}')$ (dash-dot lines 2 and $\hat{2}$), and then to the reference shrink-swell cycle of aggregated soil, $U_a(W)$ and $\hat{U}_a(\hat{W})$ (solid lines 3 and $\hat{3}$; for simplicity we take $U_s=0$, then $Y_r(W)=U_a(W)$ and $\hat{Y}_r(\hat{W})=\hat{U}_a(\hat{W})$). Dotted lines 3' and $\hat{3}'$ correspond to aggregate volumes, $U_a$ and $\hat{U}_a$ as functions of $w'$ and $\hat{w}'$, respectively. $w'$ and $\hat{w}'$ are the intra-aggregate water content at shrinkage and swelling, respectively, per unit mass of soil solids. $W$ and $\hat{W}$ are the total water content at shrinkage and swelling, respectively, per unit mass of soil solids. $w$ is intra-aggregate water content at shrinkage and swelling, respectively, per unit mass of intra-aggregate matrix solids. The other designations are obvious.

**Fig.4.** A qualitative view of the relative characteristic internal sizes of clay pores in the intra-aggregate matrix against the relative water content at shrinkage and swelling. The "relative" size means the ratio of a size to $r_{mM}$ (the maximum external clay pore size at the liquid limit); the subscript $i$ of $r_i$ corresponds to the index of the shown curves, $i=1,\ldots,6$. 1 - the maximum internal size of clay pores at shrinkage, $r_m(\zeta)/r_{mM}$ in the range $0\leq\zeta\leq\zeta_h$; 2 - the minimum internal size of clay pores at shrinkage, $r_o(\zeta)/r_{mM}$ in the range $0\leq\zeta\leq\zeta_h$; 3 - the maximum internal size of water-filled clay pores at shrinkage, $R(\zeta)/r_{mM}$; in the range $\zeta_n\leq\zeta\leq\zeta_h$ $R(\zeta)=r_m(\zeta)$; 4 - the maximum internal size of clay pores at swelling, $\hat{r}_m(\zeta)/r_{mM}$ in the range $0\leq\zeta\leq\zeta_h$; 5 - the minimum internal size of clay pores at swelling, $\hat{r}_o(\zeta)/r_{mM}$ in the range $0\leq\zeta\leq\zeta_h$; 6 - the maximum internal size of water-filled clay pores at swelling, $\hat{R}(\zeta)/r_{mM}$. $R_{min}$ and $R_m$ are the minimum and maximum sizes of non-shrinking and non-swelling clay pores in the interface layer part of the $U_i$ volume at shrinkage and swelling. $R_{min}=r_{ms}=r_m(\zeta_s)$

and $R_m=r_{mh}=r_m(\zeta_h)$. $\hat{R}_{min}$ and $\hat{R}_m$ are the minimum and maximum sizes of non-shrinking and non-swelling clay pores in the interface layer part of the $\Delta\hat{U}_i$ volume at swelling. $\hat{R}_{min}=r_{oz}=r_o(0)$ and $\hat{R}_m=r_{mz}=r_m(0)$. $\zeta_a$ is the end point of the capillary water decrease at shrinkage when only an *adsorbed* film remains; $\zeta_z$ is the shrinkage limit; $\zeta_e$ is the *end* point of water filling in the interface layer part of the $\Delta\hat{U}_i$ volume at swelling; $\zeta_n$ is the air-entry point; $\zeta_b$ is the *beginning* point of water filling in the interface layer part of the $U_i$ volume at swelling; $\zeta_s$ corresponds to the end point of the *structural* shrinkage range of the soil to which the clay contributes; $\zeta_h$ is the maximum swelling point of clay.

**Fig.5.** The qualitative view of the different shrinkage and swelling curves of an aggregated soil. $Y_r(W)$ is the reference shrinkage curve [13-15]. $Y(W)$ is the shrinkage curve with cracks (depending on sample size or layer thickness) [17]. The difference between $Y(W)$ and $Y_r(W)$ is stipulated by the development of crack volume, $U_{cr}(W)-U_s$ at shrinkage (depending on sample size or layer thickness). $\hat{Y}_r(\hat{W})$ is the reference swelling curve (Section 4.2). $\hat{Y}(\hat{W})$ is the swelling curve with cracks (depending on sample size or layer thickness; Section 5.2). The difference between $\hat{Y}(\hat{W})$ and $\hat{Y}_r(\hat{W})$ is stipulated by the crack volume, $\hat{U}_{cr}(\hat{W})-U_s$ (depending on sample size or layer thickness) that decreases at swelling from $U_{crz}-U_s=\hat{U}_{crz}-U_s$ to the $\hat{U}_{crh}-U_s$ value, that is, the specific residual volume of the non-totally closed cracks after the shrink-swell cycle (per unit mass of soil solids). $\Delta Y_h \equiv Y_h - \hat{Y}_h$ is the decrease in maximum specific swelling volume after the shrink-swell cycle of aggregated soil. $\Delta Y_h$ includes the contribution of the reference shrink-swell cycle, $\Delta Y_{rh} \equiv Y_{rh} - \hat{Y}_{rh}$ minus the contribution of the residual crack volume, $\hat{U}_{crh}-U_s$. With water content increase at the end of swelling in the range, $\hat{W}_h \leq \hat{W} \leq \hat{W}_h^*$ the residual cracks are filled in water.

**Fig.6.** White squares and circles indicate the experimental points of the shrinkage ($e^{exp}(\vartheta)$) and swelling ($\hat{e}^{exp}(\hat{\vartheta})$) curves, respectively, of soil 03 from [1] obtained on samples with initial height $h$=6.1cm. Curve 1 gives two shrinkage curves predicted with fitting and almost coinciding with each other: the reference shrinkage curve, $e_r(\vartheta)$ and shrinkage curve with cracks for sample ($h$=6.1cm), $e(\vartheta)$. Curve 3 is the predicted shrinkage curve for layer ($h$=6.1cm), $e_l(\vartheta)$. The predicted slopes in the basic shrinkage range of $e_r(\vartheta)$, $e(\vartheta)$, and $e_l(\vartheta)$ are $S$=1, 0.9972, and 0.8086, respectively. Curve 2 gives two physically predicted swelling curves that almost coincide with each other: the reference swelling curve, $\hat{e}_r(\hat{\vartheta})$ and swelling curve with cracks for sample ($h$=6.1cm), $\hat{e}(\hat{\vartheta})$. Curve 4 is the predicted swelling curve for layer ($h$=6.1cm), $\hat{e}_l(\hat{\vartheta})$. The predicted initial slopes of $\hat{e}_r(\hat{\vartheta})$, $\hat{e}(\hat{\vartheta})$, and $\hat{e}_l(\hat{\vartheta})$ are 0.7496, 0.7475, and 0.6061, respectively. The dashed line is the sample swelling curve ($h$=6.1cm) predicted with fitting to compare with $\hat{e}(\hat{\vartheta})$ (curve 2). $\hat{\vartheta}_{hs}$ and $\hat{\vartheta}_{hl}$ correspond to $\hat{\vartheta}_h^*$ (in Fig.5 it is $\hat{W}_h^*$) for sample and layer, respectively (for samples under consideration $\hat{\vartheta}_h \cong \hat{\vartheta}_{hs}$).

**Fig.7.** As in Fig.6, but for soil 04 from [1]. The predicted slopes in the basic shrinkage range of $e_r(\vartheta)$, $e(\vartheta)$, and $e_l(\vartheta)$ are $S$=1, 0.9954, and 0.7950, respectively. The predicted initial slopes of $\hat{e}_r(\hat{\vartheta})$, $\hat{e}(\hat{\vartheta})$, and $\hat{e}_l(\hat{\vartheta})$ are 0.9681, 0.9637, and 0.7696, respectively.

**Fig.8.** The predicted relative shrinkage ($v(\zeta)$) and swelling ($\hat{v}(\zeta)$) curves ($\zeta \equiv \hat{\zeta}$) of the clay that contributes to soil 03 (as an example; cf. Fig.2). The initial slope of $\hat{v}(\zeta)$ is 0.9816.

**Fig.9.** The predicted shrinkage ($U(w)$) and swelling ($\hat{U}(w)$) curves ($w \equiv \hat{w}$) of the intra-aggregate matrix of soil 04 (as an example). The initial slope of $\hat{U}(w)$ is 1.1687 dm$^3$kg$^{-1}$.

**Fig.10.** Curve 6 presents the (relative) maximum internal size of the water-filled clay pores at contributive-clay swelling, $\hat{R}(\zeta)/r_{mM}$ (the solution of Eq.(69)) for soil 03 as an example. Curves 4 and 5 present the (relative) maximum and minimum internal sizes of clay pores at swelling, $\hat{r}_m(\zeta)/r_{mM}$ (Eq.(11a) with replacement $v(\zeta) \to \hat{v}(\zeta)$) and $\hat{r}_o(\zeta)/r_{mM}$ (Eq.(11b) with replacement $v(\zeta) \to \hat{v}(\zeta)$), respectively, for soil 03. Characteristic water content values are: $\zeta_e=0.2125$, $\zeta_b=0.2525$, $\zeta_h=0.5$; characteristic pore sizes are $R_{min}/r_{mM}=0.7629$, $R_m/r_{mM}=0.8185$, $\hat{R}_{min}/r_{mM}=0.0530$, and $\hat{R}_m/r_{mM}=0.6720$ (cf. Fig.4).

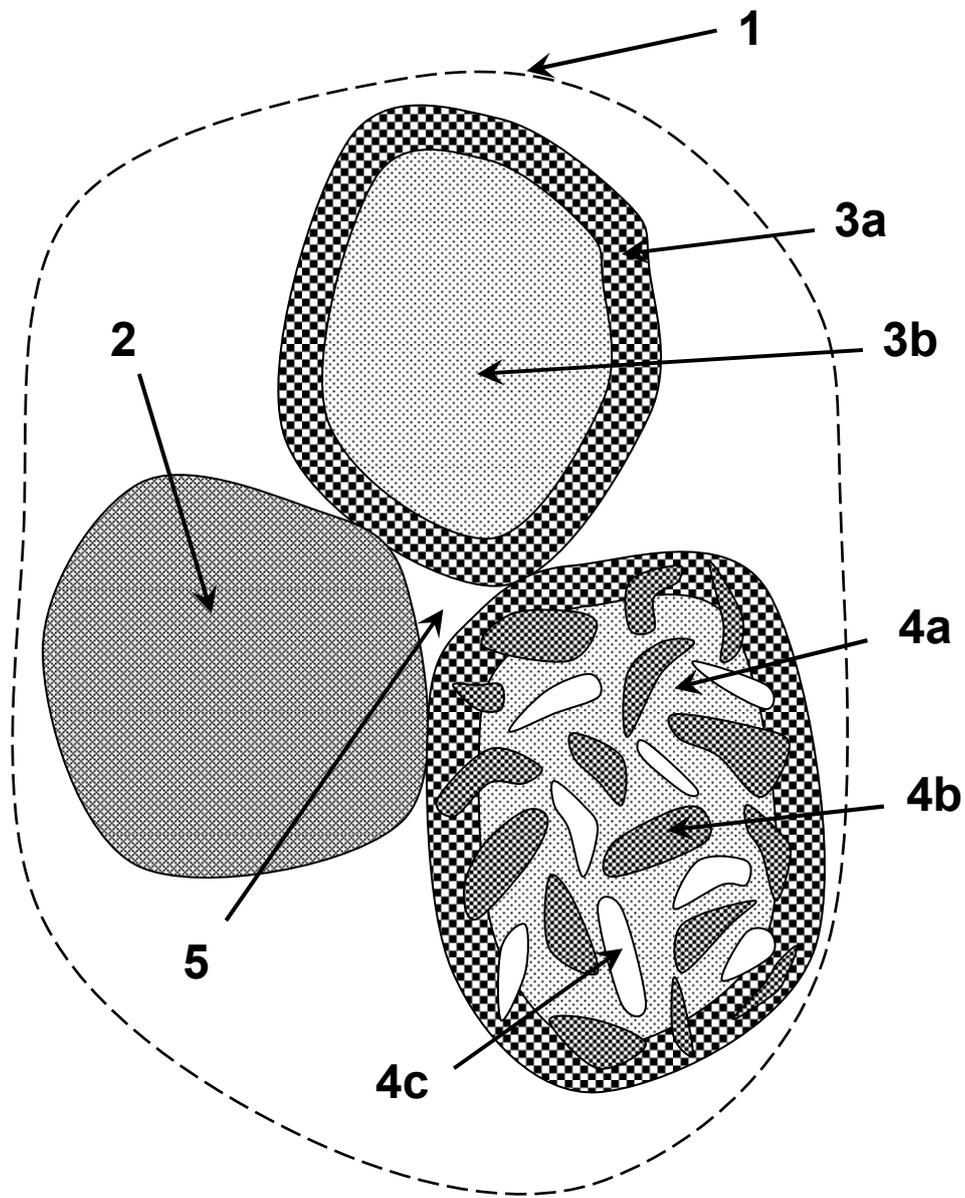

Fig.1

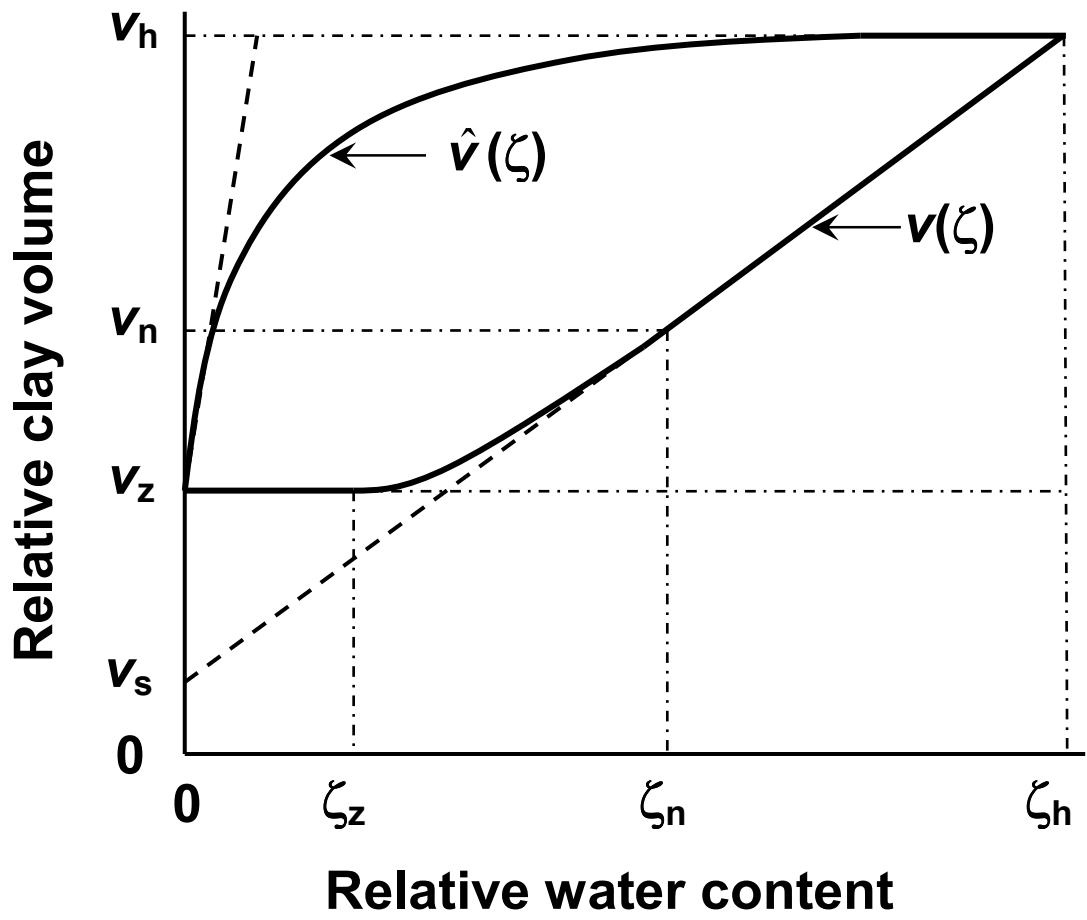

Fig.2

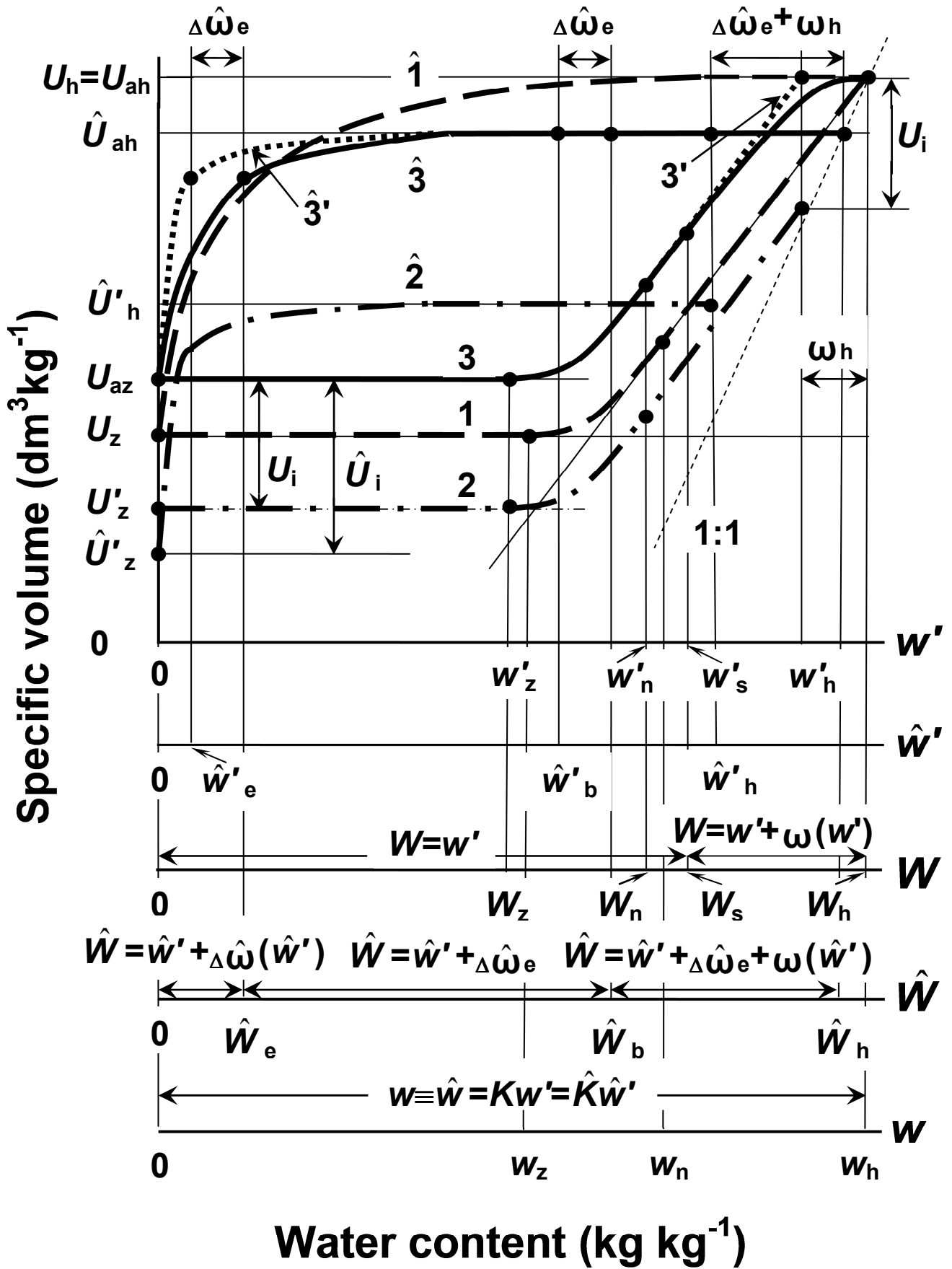

Fig.3

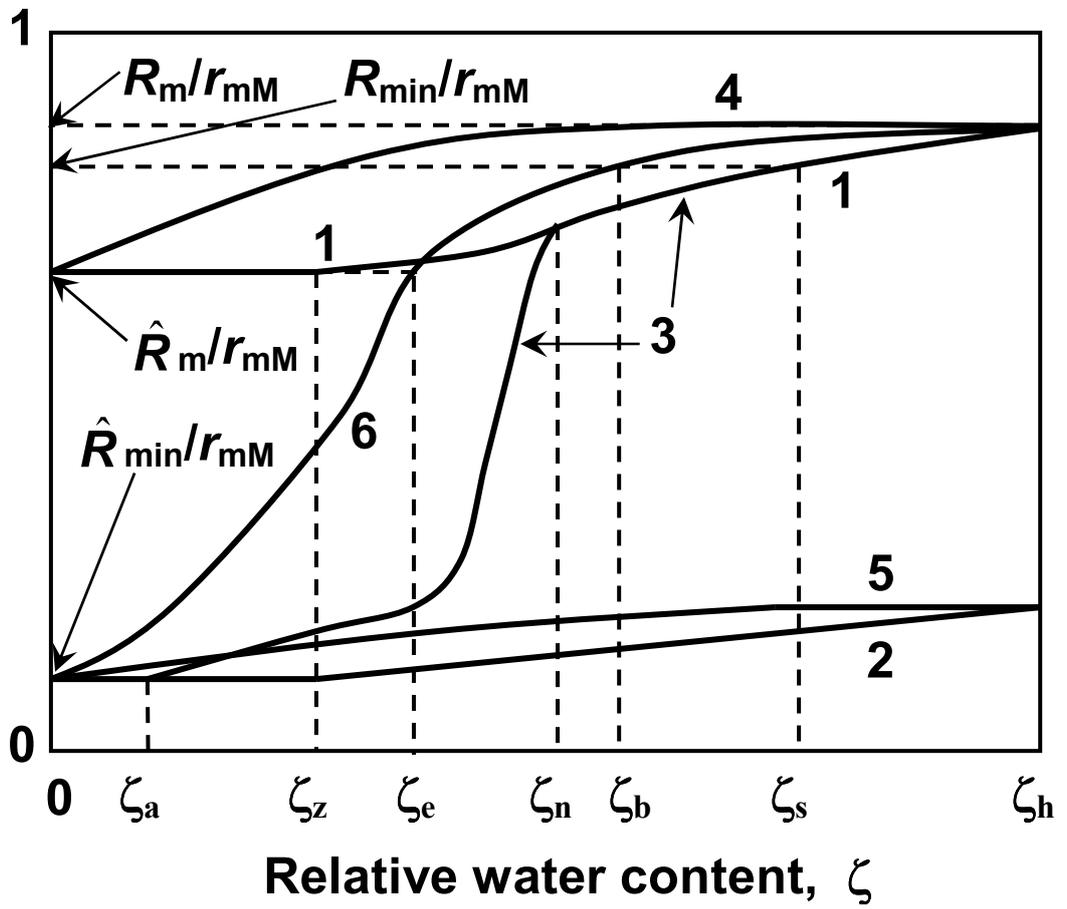

Fig.4

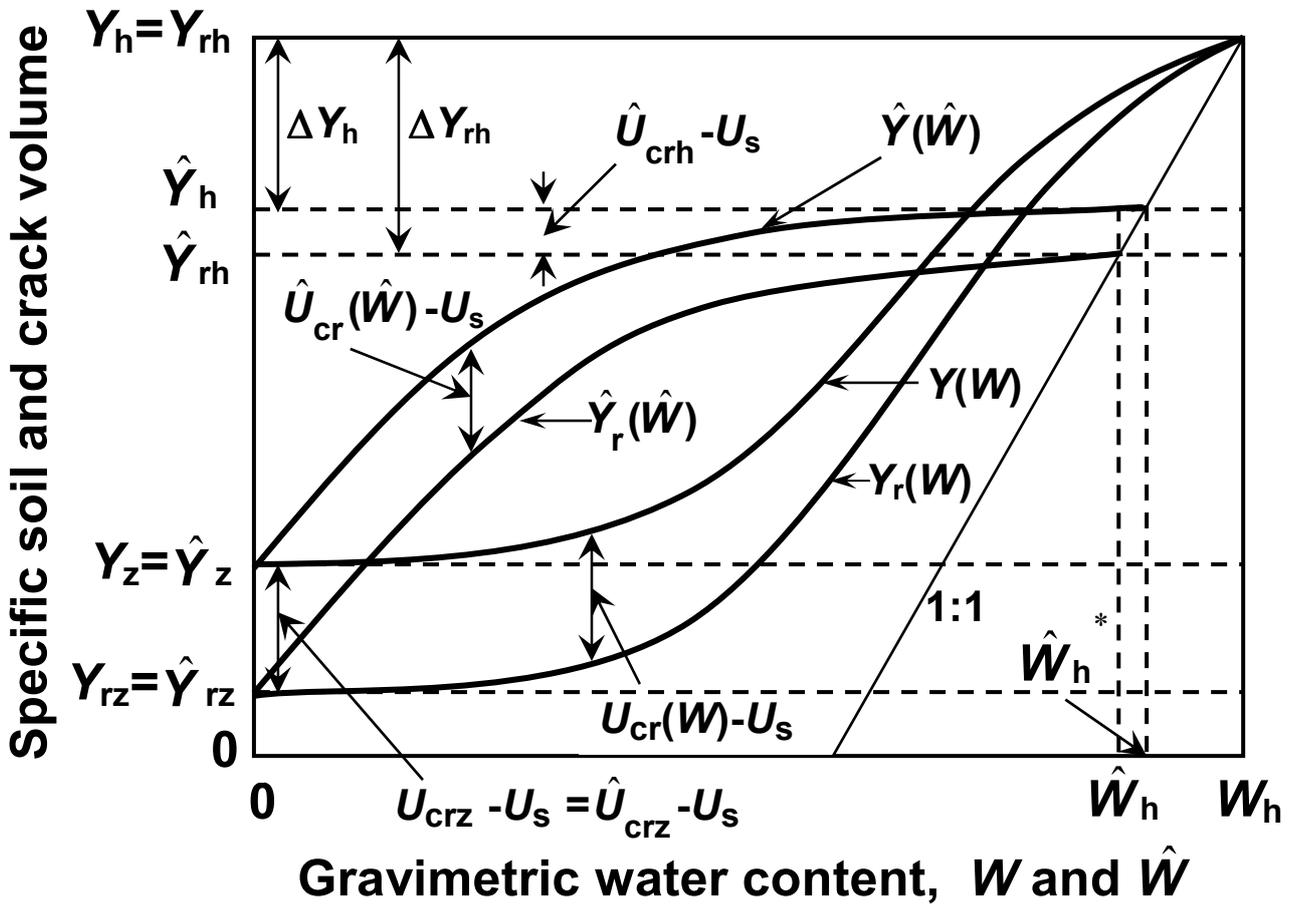

Fig.5

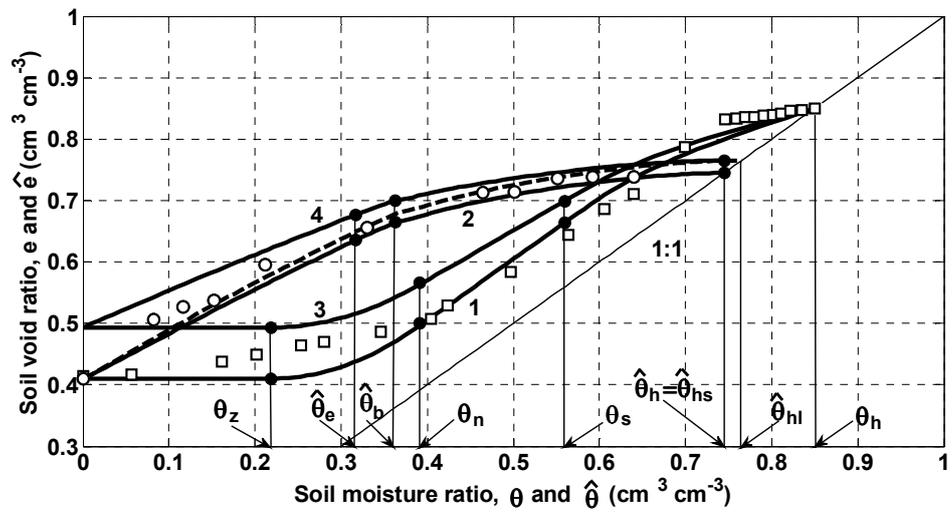

Fig.6

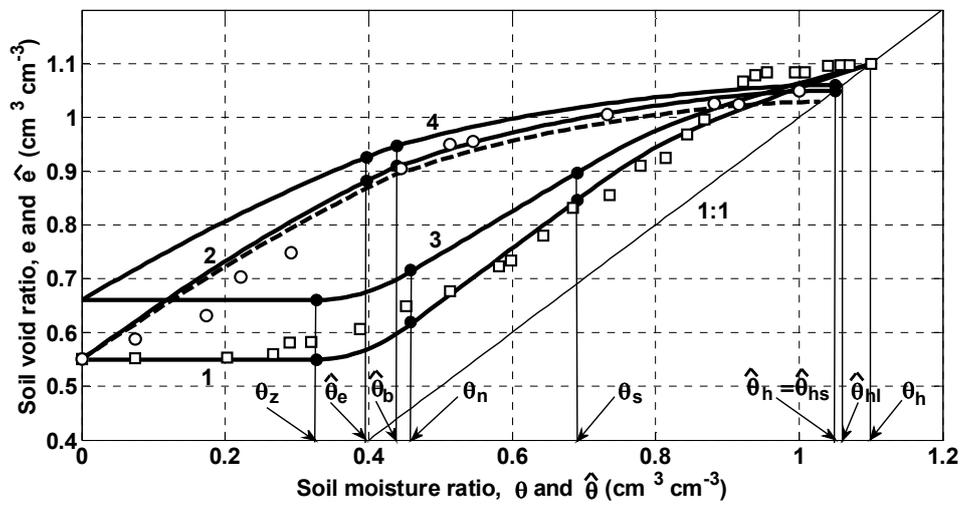

Fig.7

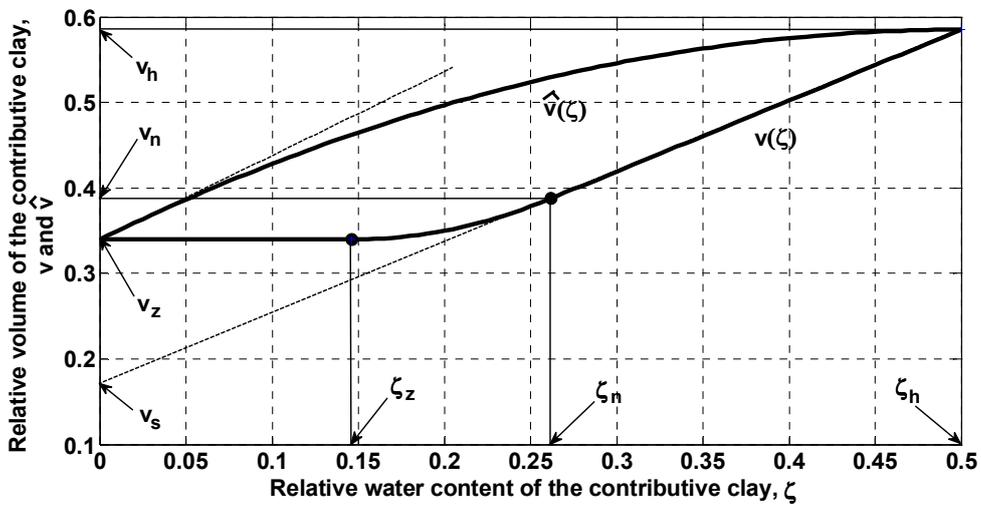

Fig.8

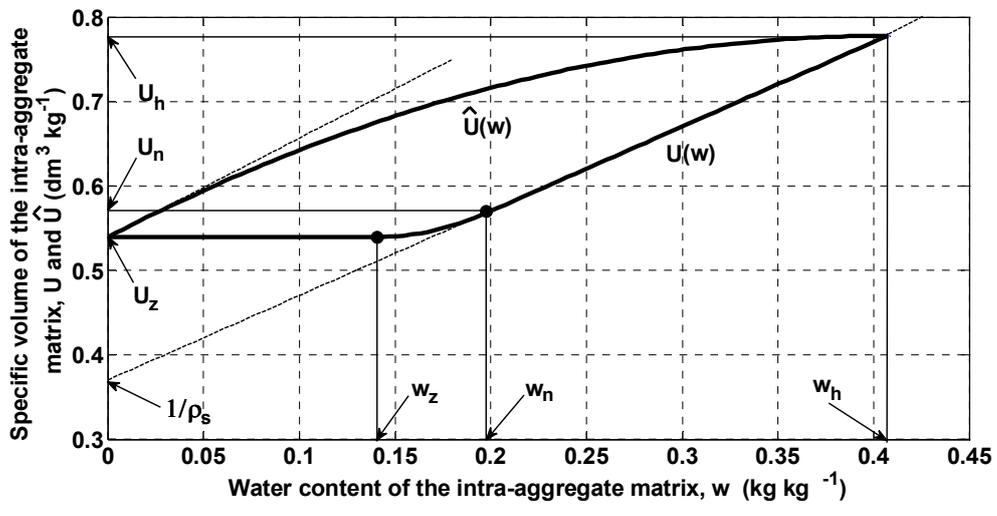

Fig.9

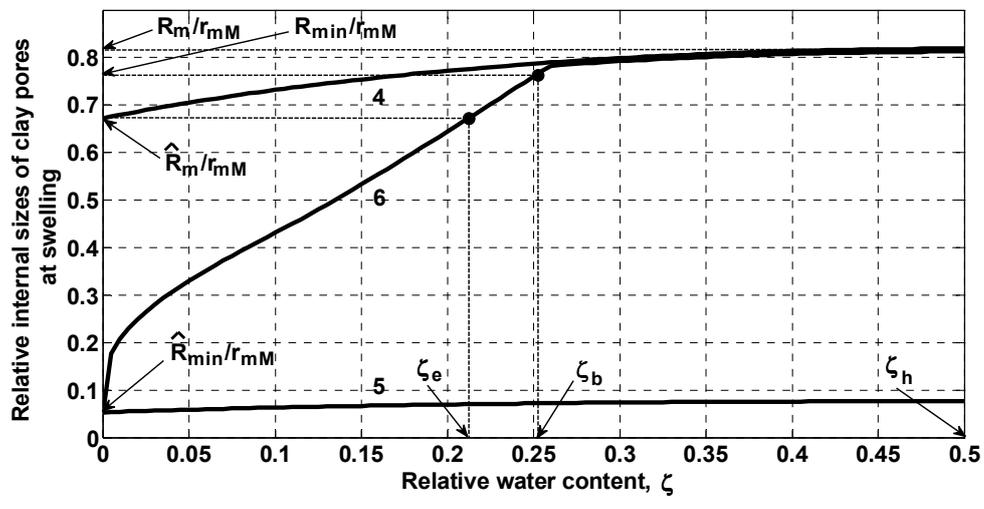

Fig.10

**Table 1.** Primary data* and input physical soil parameters for model prediction**

| Data source | Primary data | | Input parameters | | | | | | | | | |
|---|---|---|---|---|---|---|---|---|---|---|---|---|
| | $\vartheta_h$ | $e_z$ | $\rho_s$ (kg dm$^{-3}$) | $c$ | $s_1$ | $s_2$ =1-$c$-$s_1$ | $x_m$ (mm) | $X_{mz}$ (mm) | $P_z$ | $W_h$ (kg kg$^{-1}$) | $Y_z$ (dm$^3$kg$^{-1}$) | $U_{lph}$ (dm$^3$kg$^{-1}$) |
| Peng and Horn [1], soil 03 (Calcic Gleysol) | 0.85 | 0.41 | 2.7 | 0.349 | 0.577 | 0.074 | 0.072 | 1.795 | 0 | 0.3148 | 0.5222 | 0 |
| As above, soil 04 (Dystric Gleysol) | 1.10 | 0.55 | 2.7 | 0.649 | 0.151 | 0.200 | 0.120 | 1.650 | 0 | 0.4074 | 0.5741 | 0 |

* Soil moisture ratio at maximum swelling state before shrinkage ($\vartheta_h$), void ratio at shrinkage limit ($e_z$).
** Mean solid density ($\rho_s$), clay content ($c$), silt content ($s_1$), sand content ($s_2$), estimated value of the maximum sand grain size ($x_m$), estimated value of the maximum aggregate size at the shrinkage limit ($X_{mz}$), structural porosity at the shrinkage limit ($P_z$), soil water content at maximum swelling before shrinkage ($W_h$), specific soil volume at the shrinkage limit ($Y_z$), specific lacunar pore volume at maximum swelling before shrinkage ($U_{lph}$).

**Table 2.** Primary data of the decrease in maximum soil volume and model prediction for layer case (in $(\vartheta, e)$ and $(\hat{\vartheta}, \hat{e})$ variables)

| Data source | Primary data[*] | | | Model prediction in layer case[#] | |
|---|---|---|---|---|---|
| | $e_h^{exp}$ | $\hat{e}_h^{exp}$ | $\Delta e_h^{exp}$ $= e_h^{exp} - \hat{e}_h^{exp}$ | $(\Delta e_h)_l$ | $(\hat{e}_{crh} - e_s)_l$ |
| Peng and Horn [1], soil 03 (Calcic Gleysol) | 0.85 | 0.73 | 0.12 | 0.085 | 0.02 |
| As above, soil 04 (Dystric Gleysol) | 1.10 | 1.05 | 0.05 | 0.04 | 0.01 |

[*] Experimental void ratio at maximum swelling state before shrinkage ($e_h^{exp}$), experimental void ratio at maximum swelling state after shrink-swell cycle ($\hat{e}_h^{exp}$), experimental maximum-soil-volume decrease in sample case ($\Delta e_h^{exp}$).
[#] Predicted maximum-soil-volume decrease in layer case ($\Delta e_h)_l$, predicted residual crack volume in layer case ($\hat{e}_{crh} - e_s)_l$

**Table 3.** The estimated parameters of texture and structure*, cracking**, and intra-aggregate structure[#] of two soils during shrinkage as well as estimated characteristics[##] of fitting of the model shrinkage curve, $e(\vartheta)$ to shrinkage curve data

| Soil | $x_n(=X_{min})$ (mm) | $X_{mz}$ (mm) | $X_m$ (mm) | $l_{min}$ (mm) | $l_m$ (mm) | $h_*$ (cm) | $h/h_*$ | $q_s$ | $q_l$ | $k$ | $K$ | $r_e^2$ | $\sigma_e$ |
|---|---|---|---|---|---|---|---|---|---|---|---|---|---|
| Soil 03 | 0.020 | 1.795 | 1.993 | 1.006 | 145.772 | 5.3998 | 1.1297 | 0.0028 | 0.1914 | 0 | 1.1396 | 0.9819 | 0.0243 |
| Soil 04 | 0.022 | 1.650 | 1.861 | 0.9523 | 129.723 | 5.2176 | 1.1691 | 0.0046 | 0.2050 | 0 | 1.1633 | 0.9838 | 0.0272 |

* Mean size of soil solids ($x_n=0.001c+0.026s_1+(0.025+x_m/2)s_2$), minimum aggregate size ($X_{min}$), maximum aggregate size at shrinkage limit ($X_{mz}$), maximum aggregate size at maximum swelling before shrinkage ($X_m$), mean distance between the smallest ($l_{min}$) and largest ($l_m$) aggregates at maximum swelling before shrinkage.
** Critical sample/layer size at shrinkage ($h_*$), relative sample/layer size at shrinkage ($h/h_*$), sample crack factor at shrinkage ($q_s$), layer crack factor at shrinkage ($q_l$).
[#] Soil lacunar factor ($k$), aggregate/intra-aggregate mass ratio at shrinkage ($K$).
[##] Goodness of fit of $e(\vartheta)$ ($X_{mz}$ is fitting parameter) to shrinkage curve data ($r_e^2$), standard deviations of shrinkage curve data ($\sigma_e$).

**Table 4.** Intermediate soil characteristics* participating in the $k$, $K$, $q_s$, and $q_l$ calculation of two soils

| Soil | $u_s$ | $u_S$ | $u_z$ | $u_h$ | $U_h$ (dm³kg⁻¹) | $U_z$ (dm³kg⁻¹) | $U_i$ (dm³kg⁻¹) | $v_s$ | $v_z$ | $v_h$ | $f_{silt}$ | $p$ | $c*$ |
|---|---|---|---|---|---|---|---|---|---|---|---|---|---|
| Soil 03 | 0.3704 | 0.2411 | 0.4989 | 0.6852 | 0.6852 | 0.4989 | 0.0840 | 0.1703 | 0.3397 | 0.5852 | 0.8863 | 0.2225 | 0.1255 |
| Soil 04 | 0.3125 | 0.1097 | 0.4553 | 0.6563 | 0.7778 | 0.5396 | 0.1092 | 0.2278 | 0.3882 | 0.6139 | 0.4302 | 0.2875 | 0.1914 |

* Relative volume of all solids of the intra-aggregate matrix ($u_s$), relative volume of non-clay solids of the intra-aggregate matrix ($u_S$), relative volume of the intra-aggregate matrix at shrinkage limit ($u_z$), relative volume of the intra-aggregate matrix at maximum swelling ($u_h$), specific volume of the intra-aggregate matrix at maximum swelling ($U_h$), specific volume of the intra-aggregate matrix at shrinkage limit ($U_z$), contribution of the interface aggregate layer to the specific volume of soil aggregates at shrinkage ($U_i$), relative volume of the contributive-clay solids ($v_s$), relative volume of the contributive-clay matrix at shrinkage limit ($v_z$), relative volume of the contributive-clay matrix at maximum swelling ($v_h$), weight fraction of the silt grains among all silt and sand grains ($f_{silt}$), porosity of the contributive silt and sand grains at *imagined* contact ($p$), critical clay content ($c*$).

**Table 5.** Some estimated parameters of two soils during swelling after shrinkage*

| Soil | $\hat{U}_i$ (dm$^3$kg$^{-1}$) | $\hat{K}$ | $\hat{X}_{mz}$ (mm) | $\hat{X}_m$ (mm) | $\hat{K}_f \pm \Delta\hat{K}_f$ | $r_{\hat{e}}^2$ | $\sigma_{\hat{e}}$ |
|---|---|---|---|---|---|---|---|
| Soil 03 | 0.1882 | 1.4958 | 0.1360 | 0.1490 | 1.41±0.10 | 0.9854 | 0.0152 |
| Soil 04 | 0.1514 | 1.2796 | 0.2075 | 0.2316 | 1.34±0.10 | 0.9655 | 0.0311 |

* Physically estimated interface layer contribution to the specific volume of aggregates at swelling ($\hat{U}_i$), aggregate/intra-aggregate mass ratio at swelling corresponding to $\hat{U}_i$ ($\hat{K}$), maximum aggregate size after their destruction close to $\hat{W}=0$ ($\hat{X}_{mz}$) and at maximum swelling ($\hat{X}_m$), aggregate/intra-aggregate mass ratio at swelling found by fitting and its standard deviation ($\hat{K}_f \pm \Delta\hat{K}_f$), goodness of fit of $\hat{e}(\hat{\vartheta})$ to sample swelling curve data ($r_{\hat{e}}^2$), standard deviation of sample swelling curve data ($\sigma_{\hat{e}}$).